\documentclass[a4paper,12pt]{article}

\usepackage{epsfig,graphicx,amsmath,amssymb,citesort}
\def \simlt {\stackrel{<}{\sim}}
\newcommand{\gvga}{$g_{V\gamma}~$}
\newcommand{\ee}{e^{+} e^{-}}
\def\amu{a_\mu}
\def\amuh{a_\mu^{{\mathrm{had}}}}
\newcommand{\gv}{\mbox{GeV}}
\newcommand{\epm}{e^+e^-}
\newcommand{\power}[1]{\times 10^{#1}}
\newcommand{\mbo}[1]{$ #1 $}
\begin{document}

\thispagestyle{empty}

$\phantom{.}$

\hfill

\begin{center}
{\Large {\bf MesonNet Workshop on } \\
\vspace{0.75cm}
{\huge {\bf  Meson Transition Form Factors
}}}

\vspace{1cm}

{\large May 29--30, 2012 in Cracow, Poland}

\vspace{2cm}

{\it Editors:}
 E.~Czerwi\'nski (Cracow), S.~Eidelman (Novosibirsk), C.~Hanhart (J\"ulich), B.~Kubis (Bonn), A.~Kup\'s\'c
 (Uppsala), S.~Leupold (Uppsala), P.~Moskal (Cracow),\\
 and S.~Schadmand (J\"ulich)

\vspace{2.5cm}

ABSTRACT

\end{center}

\vspace{0.3cm}

\noindent
The mini-proceedings 
of the Workshop on Meson Transition Form Factors held in 
Cracow from May 29$^{\rm th}$ to 30$^{\rm th}$, 2012 
introduce the meson transition form factor
project with special emphasis on the interrelations between the various form
factors (on-shell, single off-shell, double off-shell).
Short summaries of the talks presented at the workshop follow.

\medskip\noindent
The web page of the conference, which contains all talks, can be found at
\begin{center}
{\tt\footnotesize https://sites.google.com/site/mesonnetwork/home/activities/form-factor-workshop-2012}
\end{center}

\vspace{0.5cm}

\noindent
We acknowledge the support of the European Community Research Infrastructure Integrating Activity 
``Study of Strongly Interacting Matter'' (acronym HadronPhysics3, Grant Agreement n.~283286) under the 
Seventh Framework Programme of the EU.
Furthermore, the meeting would not have been possible without the support
of the Jagiellonian University Krak\'ow.

\newpage

{$\phantom{=}$}

\vspace{0.5cm}

\tableofcontents

\newpage

\section{Introduction -- Scope of the workshop}

\addtocontents{toc}{\hspace{1cm}{\sl S.~Eidelman et al.}\par}

\vspace{5mm}

S.~Eidelman$^1$, C.~Hanhart$^2$, B.~Kubis$^3$, A.~Kup\'s\'c$^4$ 
and S.~Leupold$^4$

\vspace{5mm}

\noindent
{\footnotesize
$^1$ Budker Institute of Nuclear Physics SB RAS and Novosibirsk State University\\
$^2$ Institut f\"ur Kernphysik (Theorie) and J\"ulich Center for Hadron
Physics, Forschungszentrum J\"ulich\\
$^{3}$ HISKP (Theorie) and Bethe Center for Theoretical Physics, Universit\"at Bonn\\
$^4$ Uppsala University
}

\subsection{Motivation}

Transition form factors are an important ingredient in the detailed
understanding of the nature of mesons and their underlying quark and
gluon structure.
Lately, meson transition form factors have been widely discussed for a
number of reasons: 
\begin{itemize}
\item The field is relevant to better quantify the Standard Model value for the
anomalous magnetic moment of the muon $(g-2)$ ($a_{\mu}$).
In particular for the calculation of the hadronic light-by-light (LbL)
contribution~\cite{Hayakawa:1997rq,Bijnens:2007pz,Prades:2009tw,Jegerlehner:2009ry}, see Fig.~\ref{fig:lbl}, 
one needs to know various form factors describing the interaction of photons with hadrons.   
\begin{figure}[b]
  \centering
  \includegraphics[keepaspectratio,width=0.6\textwidth]{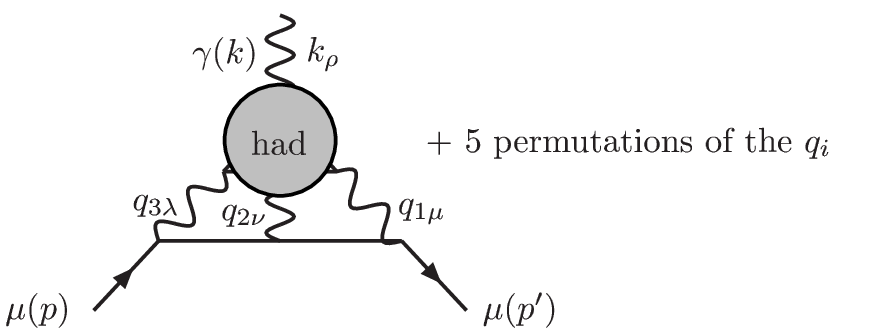}
  \caption{Light-by-light scattering contribution. 
Figure taken from Ref.~\cite{Jegerlehner:2009ry}.}
  \label{fig:lbl}
\end{figure}
\item Precise knowledge of the lepton pair mass spectra is mandatory in
a search for the quark-gluon plasma and medium modifications of 
hadron properties in heavy-ion collisions~\cite{Friman:2011zz}.
\item It is a field of hadronic physics where high-precision measurements 
are possible and theoretical calculations are therefore highly needed~\cite{Landsberg:1986fd}.
\end{itemize}
A detailed analysis of various theoretical approaches to LbL
in  Ref.~\cite{Prades:2009tw}  leads to $a_{\mu}^{\rm LbL,had}=(10.5  \pm  2.6)  \cdot
10^{-10}$. The estimation of Ref.~\cite{Jegerlehner:2009ry} leads to an
even larger uncertainty, $(11.6 \pm 3.9) \cdot 10^{-10}$. This uncertainty 
will very soon
limit the precision of the whole hadronic contribution to $a_{\mu}$. 
Note that there has been significant progress recently in computing
the leading-order hadronic contribution to $a_\mu$ on the lattice~\cite{Feng:2011zk*}, 
providing a prospect that lattice calculations can contribute to resolve the $a_mu$ discrepancy in the future. 

The leading contribution to the blob denoted by  ``had'' in 
Fig.~\ref{fig:lbl} is given by
the exchange of light hadrons that couple to two photons---since 
the anomalous magnetic moment is a static quantity, one expects 
the LbL contribution from these diagrams to scale as $1/m_{\rm had}^2$, with
$m_{\rm had}$ for the mass of the exchange hadron, leading
to a significant suppression of heavier intermediate states. 

Therefore, the contribution of the lightest state, the pion, is very
important. What enters in the pion-exchange contribution to LbL in the
$(g-2)$ is a form factor that describes the interaction of off-shell
pions with off-shell (or on-shell) photons ${\pi^0}^{*} \to
\gamma^{(*)}\gamma^{(*)}$, as defined in Refs.~\cite{Bijnens:1995xf,Hayakawa:1996ki,Jegerlehner:2007xe,Jegerlehner:book}. 
While such an off-shell pion form factor is not a physical quantity and therefore
model-dependent, any successful model needs to correctly describe the
pion transition form factor, $\pi^0 \to \gamma^{(*)}\gamma^{(*)}$,
where the pion is on-shell, but the photons can be real $(\gamma)$ or
virtual $(\gamma^*)$, with the virtuality being spacelike (electron
scattering) or timelike (dilepton production or $e^+ e^-$
annihilation). Information on the transition form factor can therefore
help to constrain the models used to evaluate hadronic LbL
scattering.

\subsection{Generalities on transition form factors}

\begin{figure}
\begin{tabular}{ll}
\begin{minipage}{0.47\linewidth}
\centering
  \includegraphics[width=0.55\textwidth,clip]{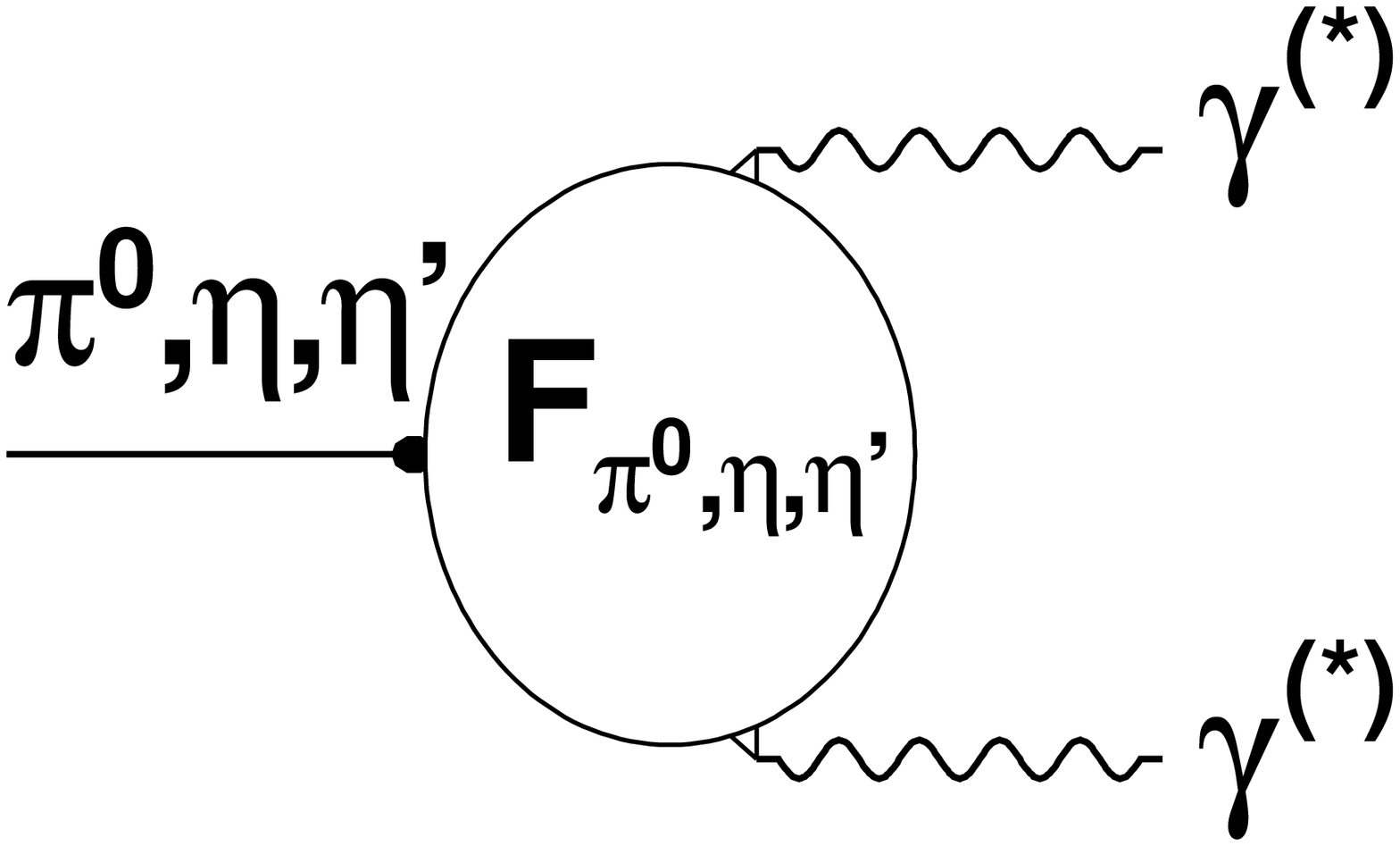}
  \caption{General $P\gamma^*\gamma^*$ vertex described by a transition form factor.}
  \label{fig:vertex} 
\end{minipage}
&
\begin{minipage}{0.47\linewidth}
\centering
  \includegraphics[keepaspectratio,width=0.55\textwidth,clip]{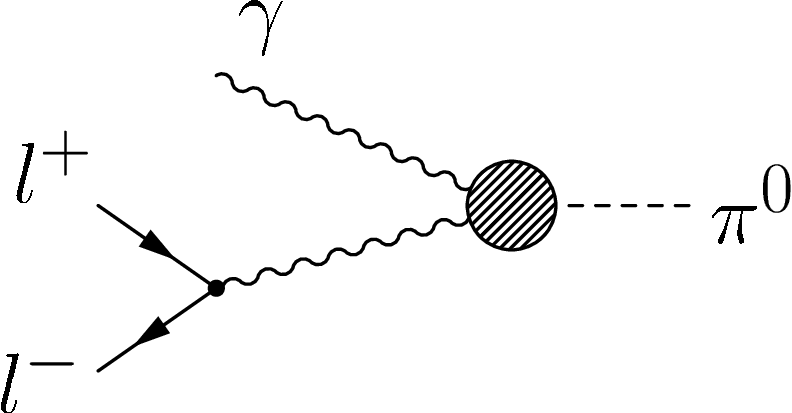}
  \caption{Pion transition form factor.}
  \label{fig:piontff}
\end{minipage}
\end{tabular}
\end{figure}

\begin{figure}[t]
\centering
\includegraphics[width=0.69\textwidth,clip]{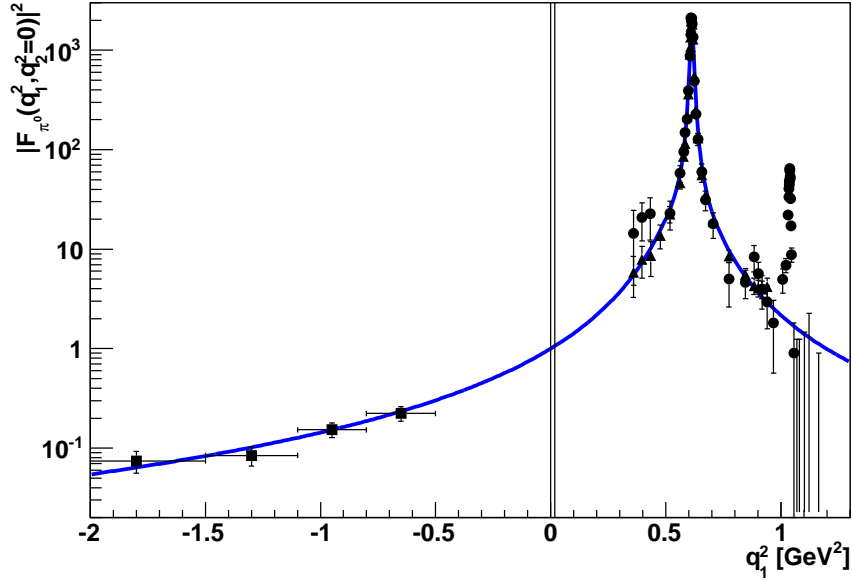} \vspace*{-2mm}
  \caption{Single off-shell $\pi^0$ meson transition form factor 
in the low $|q^2|$ region from SND~\cite{Achasov:2000ne} and 
CMD-2~\cite{Akhmetshin:2004gw} data on
the reaction $e^+e^-\to\pi^0\gamma$ 
and CELLO data on the reaction 
$e^+e^-\to e^+e^-\gamma^*\gamma^*\to  e^+e^-\pi^0$~\cite{Behrend:1990sr}.}
  \label{fig:pi0FF}
\end{figure}

\begin{figure}[t]
  \centering
  \includegraphics[keepaspectratio,width=0.18\textwidth]{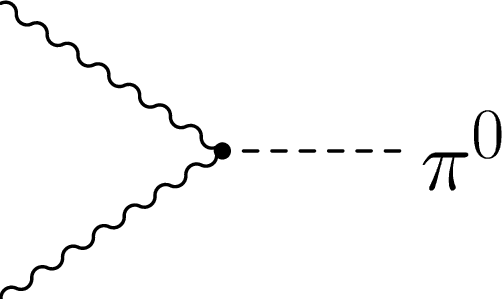} \hspace*{4em}
  \includegraphics[keepaspectratio,width=0.18\textwidth]{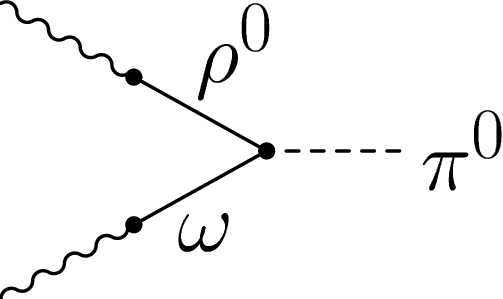} \vspace*{-2mm}
  \caption{Contributions to the pion transition form factor. 
    Wavy lines denote real or virtual photons.}
  \label{fig:ptff-contr}
\end{figure}

A transition form factor  $\mathcal{F}_P(q_1^2,q_2^2)$ is  a scalar
function of the four-momentum  transfer squared of the virtual photons
($q_{1,2}^2$)  describing the vertex in Fig.~\ref{fig:vertex}  and 
defined as
\begin{equation}
\mathcal{A}(P  \to \gamma^{*}  \gamma^{(*)}) =  q_1^{\mu} \epsilon_1^{\nu}
q_2^{\alpha}   \epsilon_2^{\beta}   \epsilon_{\mu   \nu  \alpha   \beta}
\mathcal{F}_P(q_1^2,q_2^2)
\end{equation}
and      
\begin{equation}
\frac{m_{P}^3}{64     \pi}
|\mathcal{F}_P(q_1^2=0,q_2^2=0)|^2=\Gamma(P\to\gamma\gamma).
\end{equation}
One  often uses a normalized transition form factor: 
\begin{equation}
F_P(q_1^2,q_2^2)=
\frac{\mathcal{F}_P(q_1^2,q_2^2)}{\mathcal{F}_P(q_1^2=0,q_2^2=0)} \ .
\end{equation}
For example, for the neutral pion physical processes include the following.
\begin{itemize}
\item $\pi^0 \to 2 \gamma$: this is  well described by the 
  chiral anomaly encoded in the Wess--Zumino--Witten
  action, see, e.g., Refs.~\cite{Bijnens:1988kx,Scherer:2002tk}. 
\item $\pi^0 \to \gamma e^+ e^-$ and $\pi^0 \to e^+ e^- \, e^+ e^-$: 
  as an illustration, see Fig.~\ref{fig:piontff}.
\item $e^+ e^- \to \pi^0 e^+ e^-$: here two virtual photons can 
fuse ``in flight'' to form the pion.\footnote{In principle, 
this process interferes at the amplitude level with 
$e^+ e^-$ annihilation and successive emission of
a pion and virtual photon; dominance of the virtual-photon-fusion process can be ensured by appropriate kinematical cuts.} 
\item Another process involving the pion transition 
form factor is a very rare direct dilepton
decay $\pi^0 \to e^+ e^-$. Two photons emitted in this process  
convert to a dilepton by lepton exchange, see 
Fig.~34c in Ref.~\cite{Landsberg:1986fd}. 
The pion transition form factor enters the corresponding ``QED loop''; 
see, e.g., Ref.~\cite{Bijnens:1999jp}.
\end{itemize}
Available data on $|F_{\pi^0}(q^2,0)|$ for low $|q^2|$ values are presented in Fig.~\ref{fig:pi0FF}.
For theoretical calculations of the pion form factor see, 
e.g., Refs.~\cite{Faessler:1999de,Harada:2003jx,carla-dipl,Petri} 
and references therein. The models either use strict vector meson dominance~\cite{sakurai} 
(right diagram in Fig.~\ref{fig:ptff-contr}), 
or, as, e.g., in Ref.~\cite{carla-dipl}, 
include point interactions in addition (the left diagram in Fig.~\ref{fig:ptff-contr}). 
For a review on vector mesons and their interactions, see also Ref.~\cite{meissner:physrept}.
In Ref.~\cite{Hayakawa:1997rq} the two-photon data
on the production of pseudoscalar mesons 
($\pi^0$, $\eta$, $\eta'$)~\cite{Gronberg:1997f}
was used to model
the transition form factors needed in the evaluation of  $a_{\mu}^{\rm LbL,had}$.

\begin{table}[t]
\renewcommand{\arraystretch}{1.3}
\addtocounter{footnote}{1}
\begin{center}
\begin{tabular}{lcccc}
\hline
Mode & ${\cal B}^{\rm exp}$ & Ref. & ${\cal B}^{\rm th}$ & Ref.  \\
\hline
 $\pi^0 \to \gamma\gamma$ & $(98.823 \pm 0.034)\%$ & \cite{PDG} & -- & -- \\
 $\pi^0 \to e^+e^-\gamma$ & $(1.174 \pm 0.035)$\% & \cite{PDG} & $(1.182 \pm 0.003)$\% & \cite{joseph} \\ 
 $\pi^0 \to e^+e^-e^+e^-$ & $(3.34 \pm 0.16) \cdot 10^{-5}$ & \cite{PDG} & $3.39\cdot 10^{-5}$ & \cite{Petri} \\ 
 $\pi^0 \to e^+e^-$ & $( 7.48 \pm 0.29 \pm 0.25 ) \cdot 10^{-8}$  & \cite{ktev}\protect\footnotemark[\value{footnote}]\hspace*{-1.7mm} &
 $(6.33 \pm 0.19) \cdot 10^{-8}$ & \cite{Ametller:1993we} \\
&&& $(6.1 \pm 0.3) \cdot 10^{-8}$ & \cite{Knecht:1999gb} \\ 
&&& $(6.2 \pm 0.1) \cdot 10^{-8}$ & \cite{dorokhov} \\ 
\hline
\end{tabular}
\end{center} \vspace{-3.5mm}
\caption{\label{tab:pi0} Branching fractions of $\pi^0$ radiative and leptonic decays.
See Ref.~\cite{Petri} for further theoretical references.} 
\end{table} 
In Table~\ref{tab:pi0} we list the information on the 
branching fractions of $\pi^0$ decays together with the
corresponding theoretical predictions. The branching ratios
largely follow the naive scaling as $1:\alpha_{\rm QED}:\alpha_{\rm QED}^2:\alpha_{\rm QED}^2/(4\pi)^2$,
where the factor $(4\pi)^2$ is present since in the Standard Model
the leading contribution to $\pi^0\to e^+e^-$ appears at one loop. 
Note that between the most accurate calculation for $\pi^0\to e^+e^-$
and the corresponding experimental value there is a more 
than $3\sigma$ discrepancy.
For asymptotically large virtualities there are QCD constraints on 
the pion transition form factor, 
see, e.g., Refs.~\cite{Bijnens:1999jp,Brodsky:1981rp,Brodsky:2011yv}.
These might be at odds with recent experimental results from 
BaBar~\cite{Aubert:2009mc}; see, however, 
also the recent Belle results~\cite{Uehara:2012ag}.
\footnotetext[\value{footnote}]{The value
  given in Ref.~\cite{PDG}, $(6.46 \pm 0.33) \cdot 10^{-8}$, is not corrected
for final state radiation.} 

Due to the approximate SU(3) flavor symmetry, the pion transition form factor 
is closely related to the corresponding transition form factor of the 
$\eta$ meson and, via $\eta$-$\eta'$ mixing, also to the
transition form factor of the $\eta'$. In fact, the whole discussion of 
$\eta$-$\eta'$ mixing, interesting in its own right, is strongly 
based on these transition form 
factors~\cite{Ametller:1991jv,Kaiser:1998ds,Escribano:2005qq}.

\begin{figure}[t]
\centering
\includegraphics[width=0.7\textwidth]{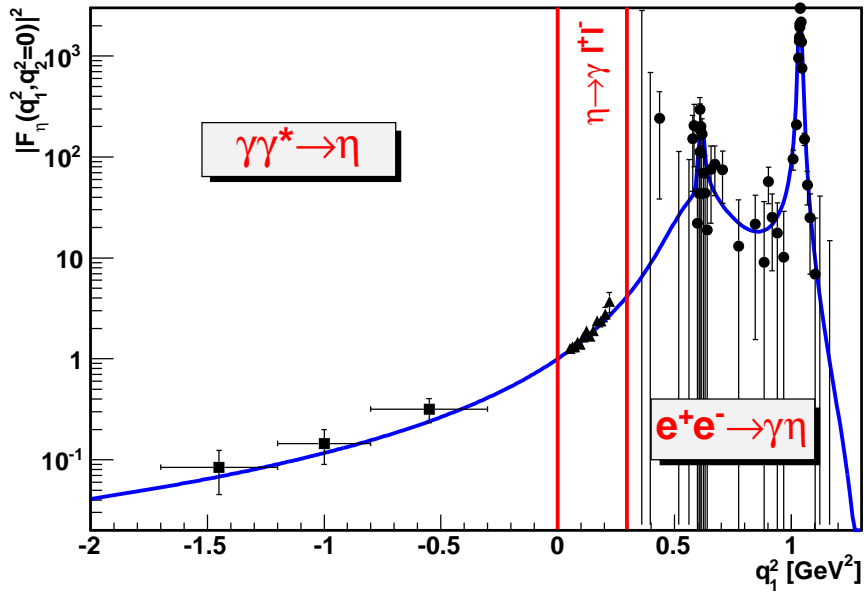} 
\caption{Single off-shell  $\eta$ meson transition form factor 
from NA60 data on
$\eta\to\gamma\mu^+\mu^-$ decay~\cite{Arnaldi:2009wb}; 
from SND~\cite{Achasov:2000ne} and CMD-2~\cite{Akhmetshin:2004gw} data on
the reaction $e^+e^-\to\eta\gamma$ reaction, and CELLO data on the reaction 
$e^+e^-\to e^+e^-\gamma^*\gamma^*\to  e^+e^-\eta$~\cite{Behrend:1990sr}.
\label{fig:etaFF}}
\end{figure}

Both processes $\pi^0 \to \gamma \gamma^*$ and $\eta \to \gamma \gamma^*$ 
can be described by vector meson
dominance~\cite{Landsberg:1986fd}. For the $\eta$ this is illustrated 
with the available low-$q^2$ data on  
$|F_{\eta}(q^2,0)|$ in Fig.~\ref{fig:etaFF}. 
Note that this does not necessarily mean that the double-virtual 
processes $\pi^0/\eta \to \gamma^* \gamma^*$ would be also well described
by vector meson dominance. Indeed, some theories 
show deviations~\cite{Borasoy:2003yb,carla-dipl}.
\begin{table}
\renewcommand{\arraystretch}{1.3}
\addtocounter{footnote}{1}
\begin{center}
\begin{tabular}{lcccc}
\hline
Mode & ${\cal B}^{\rm exp}$ & Ref. & ${\cal B}^{\rm th}$ & Ref.  \\
\hline
 $\eta \to \gamma\gamma$ & $(39.31 \pm 0.20)\%$ & \cite{PDG} & -- & -- \\
 $\eta \to e^+e^-$ & $< 5.6 \cdot 10^{-6}$ & \cite{HADES:2011} &
 $(4.5 \pm 0.2) \cdot 10^{-9}$ &  \cite{Knecht:1999gb} \\
&&& $ 5.24\cdot 10^{-9}$ & \cite{dorokhov} \\ 
 $\eta \to \mu^+\mu^-$ & $(5.8 \pm 0.8) \cdot 10^{-6}$ & \cite{PDG} &
 $(5.5 \pm 0.8) \cdot 10^{-6}$ &  \cite{Knecht:1999gb} \\
&&& $4.64\cdot10^{-6}$ &  \cite{dorokhov} \\ 
 $\eta \to e^+e^-\gamma$ & $(6.9 \pm 0.4) \cdot 10^{-3}$ & \cite{PDG} & $6.5\cdot 10^{-3}$ & \cite{Petri}  \\
 $\eta \to \mu^+\mu^-\gamma$ & $(3.1 \pm 0.4) \cdot 10^{-4}$ & \cite{Dzhelyadin:1980kh} & $3.0\cdot 10^{-3}$ &  \cite{Petri} \\  
 $\eta \to e^+e^-e^+e^-$ & $(2.4\pm0.2\pm0.1) \cdot 10^{-5}$ & \cite{KLOE2:2011aa} & $2.67\cdot 10^{-5}$ & \cite{Petri} \\
 $\eta \to e^+e^-\mu^+\mu^-$ & $< 1.6  \cdot 10^{-4}$ & \cite{Berlowski:2008zz} & $(2.2\pm0.1)\cdot 10^{-6}$ &  \cite{Petri} \\
$\eta \to \mu^+\mu^-\mu^+\mu^-$ & $< 3.6  \cdot 10^{-4}$ & \cite{Berlowski:2008zz}\protect\footnotemark[\value{footnote}]\hspace*{-1.7mm} & $(3.8\pm0.1)\cdot 10^{-9}$ &  \cite{Petri} \\  
\hline
\end{tabular}
\end{center} \vspace{-3.5mm}
\caption{\label{tab:eta} Branching fractions of $\eta$ radiative and leptonic decays.
Ref.~\cite{Petri} serves as an example for predictions of VMD-inspired models;
see references therein for further literature.}
\end{table}
\footnotetext[\value{footnote}]{Since the authors do not distinguish between the 
$\mu^+\mu^-\mu^+\mu^-$ and $\mu^+\mu^-\pi^+\pi^-$ final states, 
the upper limit is for  a sum of the branching fractions for the two modes.}\noindent
In Tables~\ref{tab:eta} and \ref{tab:etapr} we show the branching fractions
of the $\eta$ and $\eta^{\prime}$ non-hadronic decays. 

\begin{table}[t]
\renewcommand{\arraystretch}{1.3}
\begin{center}
\begin{tabular}{lcccc}
\hline
Mode & ${\cal B}^{\rm exp}$ & Ref. & ${\cal B}^{\rm th}$ & Ref.  \\
\hline
$\eta^{\prime} \to \gamma\gamma$ & $(2.18 \pm 0.08)\%$ & \cite{PDG} & -- & -- \\
 $\eta^{\prime} \to e^+e^-$ & $< 2.1 \cdot 10^{-7}$ & \cite{Briere:1999bp} &
 $1.86 \cdot 10^{-10}$ & \cite{dorokhov} \\
 $\eta^{\prime} \to \mu^+\mu^-$ & -- & -- &
 $1.30 \cdot 10^{-7}$  & \cite{dorokhov} \\ 
 $\eta^{\prime} \to e^+e^-\gamma$ & $< 9 \cdot 10^{-4}$ & \cite{vor88} & $(4.4\pm0.2)\cdot 10^{-4}$ & \cite{Petri} \\
 $\eta^{\prime} \to \mu^+\mu^-\gamma$ & $(1.07 \pm 0.26) \cdot 10^{-4}$ & \cite{vic80} & $(0.90\pm0.06)\cdot 10^{-4}$ &  \cite{Petri} \\  
 $\eta^{\prime} \to e^+e^-e^+e^-$ & -- & -- & $(2.1\pm0.1)\cdot10^{-6}$ & \cite{Petri} \\
 $\eta^{\prime} \to e^+e^-\mu^+\mu^-$ & -- & -- & $(7.6\pm0.6)\cdot10^{-7}$ &  \cite{Petri} \\
$\eta^{\prime} \to \mu^+\mu^-\mu^+\mu^-$ & -- & -- & $(2.1\pm0.2)\cdot10^{-8}$ &  \cite{Petri} \\  
\hline
\end{tabular}
\end{center} \vspace{-3mm}
\caption{\label{tab:etapr} Branching fractions of $\eta^{\prime}$ radiative and leptonic decays.
 See Ref.~\cite{Petri} for further theoretical references.} 
\end{table}

\begin{sloppypar}
Since vector mesons are relevant as intermediate states for the transition 
form factors of pseudoscalar mesons,
there is another source of information, namely the transition form factors 
of vector mesons to pseudoscalar mesons.
These involve processes like
\begin{itemize}
\item $\omega \to \pi^0 l^+ l^-$ where $l=e,\,\mu$; 
see Fig.~\ref{fig:tffomegapi} for illustration.
\begin{figure}[t]
  \centering
  \includegraphics[keepaspectratio,width=0.28\textwidth]{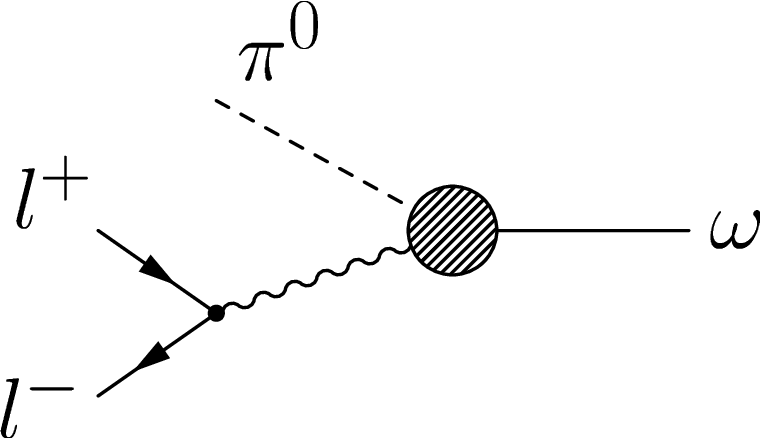}
  \hspace*{2cm}
  \includegraphics[keepaspectratio,width=0.28\textwidth]{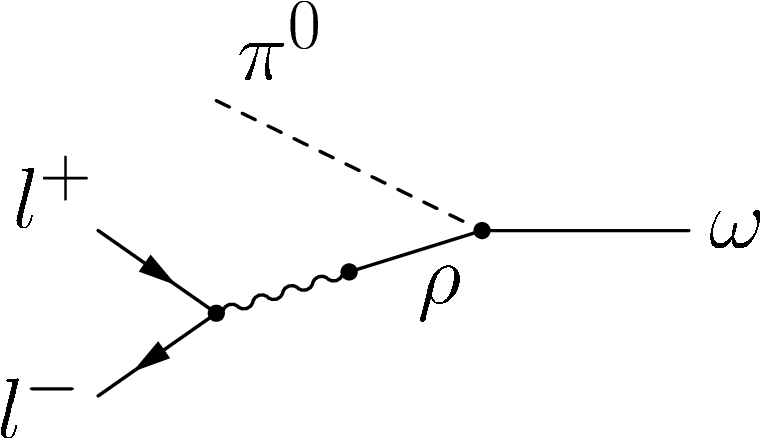}   
  \caption{Left: transition form factor of omega to pion.  
           Right: tree-level contribution to the omega transition form factor.}
  \label{fig:tffomegapi}
\end{figure}
This process shows a dramatic deviation from the vector meson dominance 
picture~\cite{Arnaldi:2009wb,Usai:2011zza}, 
see the data in Fig.~\ref{fig:omegaFF}. 
\begin{figure}[t]
\centering
\includegraphics[width=0.7\textwidth]{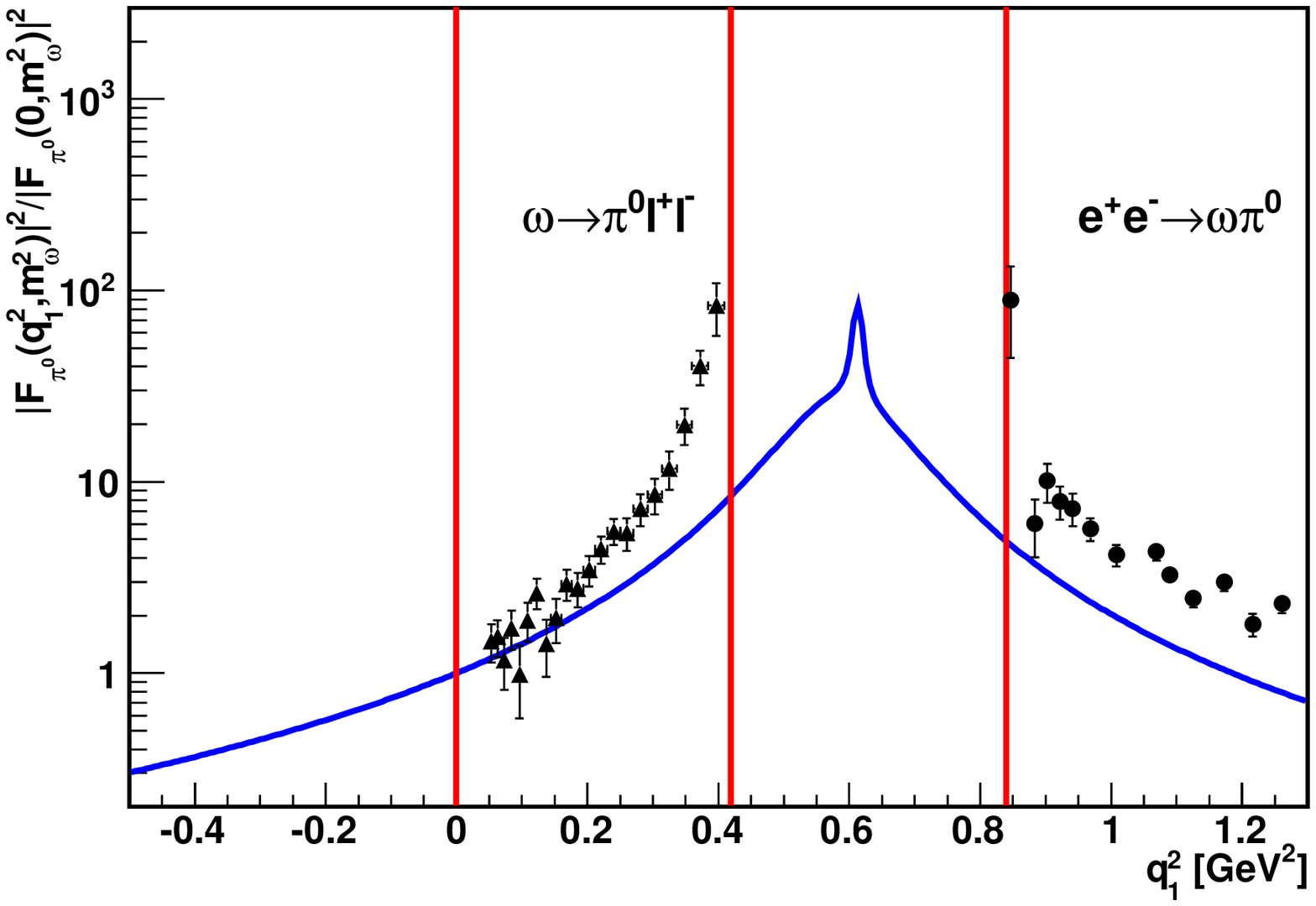}
\caption{Data on the $\omega$ transition form factor 
$|F_{\pi^0}(q^2_1,M_\omega^2)|^2/|F_{\pi^0}(0,M_\omega^2)|^2$ from 
NA60 data on $\omega\to\pi^0\mu^+\mu^-$ decay~\cite{Arnaldi:2009wb} 
and from SND~\cite{Achasov:2000wy}, CMD-2~\cite{Akhmetshin:2003ag}, 
and KLOE~\cite{Ambrosino:2008gb} experiments on the 
$e^+e^-\to\omega\pi^0$ reaction. \label{fig:omegaFF}}
\end{figure}
\item $e^+e^- \to \pi^0 \omega$: previously studied in Novosibirsk with 
SND~\cite{Achasov:2000wy} and CMD-2~\cite{Akhmetshin:2003ag}.
\item $\phi$ instead of $\omega$ in the previous processes 
and/or $\eta$ instead of $\pi^0$ (in part measured): it would be important to
clarify whether also in these processes the drastic deviation from 
vector meson dominance
seen in $\omega \to \pi^0 \mu^+ \mu^-$ shows up. Of particular importance is
$\phi \to \pi^0 l^+ l^-$, where the peak mass of the $\rho$ meson 
is in the kinematically
allowed region~\cite{Schneider:2012ez}.
\item $\eta' \to \omega \gamma$ (measured) and 
$\eta' \to \omega e^+e^-$ (not measured). 
\end{itemize}
The measured branching ratios for these vector-meson-conversion 
decays are collected
in Tables~\ref{tab:omega} and \ref{tab:phi}, compared to 
theoretical predictions.
Some of these vector-to-pseudoscalar transition form factors have been 
calculated in Refs.~\cite{Harada:2003jx,Schneider:2012ez,Bando:1993cu,Klingl,Terschluesen:2010ik,Terschlusen:2012xw}. 
The relevant diagram in the vector-meson-dominance picture is 
depicted in Fig.~\ref{fig:tffomegapi}~(right). 
Note the intimate relation of this diagram with  
Fig.~\ref{fig:ptff-contr}~(right).
\end{sloppypar}

\begin{table}[t]
\renewcommand{\arraystretch}{1.3}
\begin{center}
\begin{tabular}{lcccc}
\hline
Mode & ${\cal B}^{\rm exp}$ & Ref. & ${\cal B}^{\rm th}$ & Ref.  \\
\hline
 $\omega \to \pi^0 e^+e^-$ & $(7.7 \pm 0.6)\cdot 10^{-4}$ & \cite{PDG} & $(8.1\pm0.1) \cdot 10^{-4}$ & \cite{Terschluesen:2010ik} \\
&&& $(7.6 \ldots 8.1) \cdot 10^{-4}$ & \cite{Schneider:2012ez} \\
$\omega \to \pi^0 \mu^+\mu^-$ & $(1.73 \pm 0.25 \pm 0.14)\cdot 10^{-4}$ & \cite{Arnaldi:2009wb} & $(1.16\pm0.07) \cdot 10^{-4}$ & \cite{Terschluesen:2010ik} \\
 & $(0.96 \pm 0.23)\cdot 10^{-4}$ & \cite{Dzhelyadin:1980tj} & $(0.94 \ldots 1.00) \cdot 10^{-4}$ & \cite{Schneider:2012ez} \\
 $\omega \to \eta e^+e^-$ & $< 1.1 \cdot 10^{-5}$ & \cite{Akhmetshin:2005vy} & $(3.20 \pm 0.10)\cdot 10^{-9}$  & \cite{Terschluesen:2010ik} \\
 $\omega \to \eta \mu^+\mu^-$ & -- & -- & $(1.00 \pm 0.00)\cdot 10^{-9}$ & \cite{Terschluesen:2010ik} \\
\hline
\end{tabular}
\end{center} \vspace{-3mm}
\caption{\label{tab:omega} Branching fractions of $\omega$ conversion decays.}
\end{table}

\begin{table}[t]
\renewcommand{\arraystretch}{1.3}
\begin{center}
\begin{tabular}{lcccc}
\hline
Mode & ${\cal B}^{\rm exp}$ & Ref. & ${\cal B}^{\rm th}$ & Ref.  \\
\hline
 $\phi \to \pi^0 e^+e^-$ & $(1.12 \pm 0.28)\cdot 10^{-5}$ & \cite{PDG} & $(1.39\ldots1.53)\cdot 10^{-5}$ & \cite{Schneider:2012ez} \\
 $\phi \to \pi^0 \mu^+\mu^-$ & -- & -- & $(3.7\ldots4.1)\cdot 10^{-6}$ & \cite{Schneider:2012ez} \\
 $\phi \to \eta e^+e^-$ & $(1.15 \pm 0.10)\cdot 10^{-4}$ & \cite{PDG} & $(1.09 \pm 0.06)\cdot 10^{-4}$ & \cite{Terschluesen:2010ik} \\
$\phi \to \eta \mu^+\mu^-$ & $< 9.4 \cdot 10^{-6}$ & \cite{Akhmetshin:2000bw} & $(6.44 \pm 0.69)\cdot 10^{-6}$ & \cite{Terschluesen:2010ik} \\
\hline
\end{tabular}
\end{center} \vspace{-3mm}
\caption{\label{tab:phi} Branching fractions of $\phi$ conversion decays.}
\end{table}

To clarify the relation between the vector-meson-conversion and 
the pseudoscalar transition form factors
a little further, let us consider in particular the pion 
transition form factor.
It can be characterized in the following way:
because of isospin and G-parity, one of the virtual photons 
couples to an isovector~($v$) 
and the other to an isoscalar~($s$) state. 
Therefore, the pion transition form factor can be written as
\begin{eqnarray}
  \label{eq:fdecisiv}
  F_{\pi^0}(q_1^2,q_2^2) = F_{vs}(q_1^2,q_2^2) + F_{sv}(q_1^2,q_2^2) \,.
\end{eqnarray}
In $F_{ij}(q_1^2,q_2^2)$ the first/second index refers to the 
photon with momentum $q_1$/$q_2$.
If $\sqrt{q_2^2}$ is close to a resonance mass, $M_\omega$ or $M_\phi$, 
one can approximately neglect 
$F_{sv}(q_1^2,q_2^2)$---provided $\sqrt{q_1^2}$ is not also close to a 
resonance mass. Using the Breit--Wigner
formula~\cite{PDG}, the quantity $F_{vs}(q_1^2,q_2^2)$ and therefore 
$F_{\pi^0}$ can be approximated by
\begin{eqnarray}
  \label{eq:approxfvs}
  F_{\pi^0}(q_1^2,q_2^2) \approx f_{V \to \pi}(q_1^2) \, \frac{1}{q_2^2 - M_V^2 + i M_V \, \Gamma_{\rm tot}} \, g_{V\gamma}
  \qquad \mbox{for $q_2^2 \approx M_V^2$,}
\end{eqnarray}
where $f_{V \to \pi}(q_1^2)$ is an appropriately normalized form factor 
of the transition $V \to \pi^0 \, \gamma^*$
and $\Gamma_{\rm tot}$ the total width of the isoscalar vector meson $V$. 
This yields
\begin{eqnarray}
  \label{eq:approxfvsfin}
  \frac{F_{\pi^0}(q_1^2,M_V^2)}{F_{\pi^0}(0,M_V^2)} 
  \approx \frac{f_{V \to \pi}(q_1^2)}{f_{V \to \pi}(0)} = F_{V \to \pi}(q_1^2)  \,,
\end{eqnarray}
where the vector-to-pion transition form factor $F_{V \to \pi}$ 
is normalized to 1 at the photon point.

A transition of the omega meson to the pion is given by the 
two-dimensional $\pi^0$ form factor as:
$|F_{\pi^0}(q^2_1,M_\omega^2)|^2/|F_{\pi^0}(0,M_\omega^2)|^2$.  It was
measured       in       $\omega\to\pi^0\ell^+\ell^-$      as well as 
in the reaction
$e^+e^-\to\omega\pi^0$.   The   results   are   shown   in
Fig.~\ref{fig:omegaFF} together with naive VMD predictions. 
Note that the region around $q_1^2 \approx M_\omega^2$
is just included for completeness. There, 
Eqs.~\eqref{eq:approxfvs} and \eqref{eq:approxfvsfin} 
do not provide a good approximation.

\begin{figure}[t]
\centering
\includegraphics[width=0.65\textwidth]{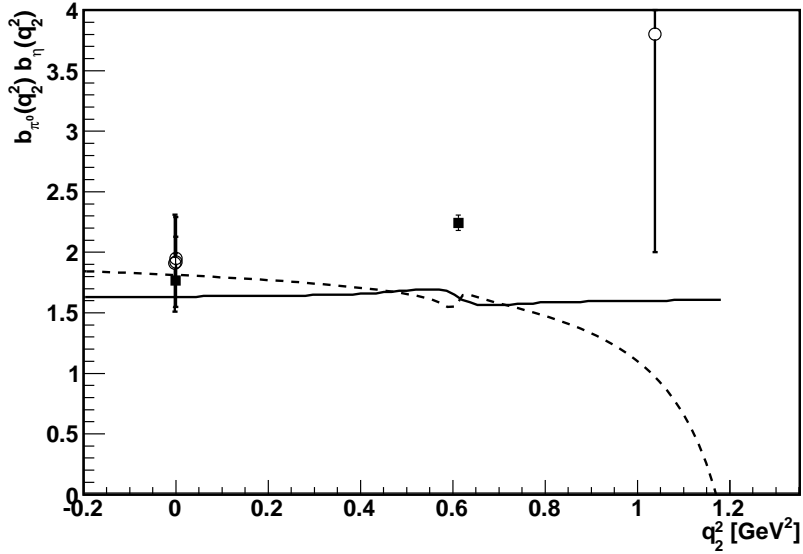}
\caption{Slopes $b_{\pi^0}(q_2^2)$ and $b_\eta(q_2^2)$ in naive VMD (solid and dashed lines, respectively), compared to
experimental data from Table~\ref{tab:lambdEx} (filled squares and open circles, respectively).
\label{fig:slopes}}
\end{figure}

A quantity often considered in the context of transition form factors 
is the form factor slope
\begin{equation}
b_P=\left.\frac{\partial \ln F(q^2,0)}{\partial q^2}\right|_{q^2=0} ~,
\end{equation}
which can be generalized to
\begin{equation}
b_P(q_2^2)=\left.\frac{\partial \ln |F(q_1^2,q_2^2)|}{\partial q_1^2}\right|_{q_1^2=0}   \,.
\end{equation}
Experimental data on the slope parameters for the $\pi^0$ transition 
form factor (both for $q^2=0$ and $q^2=M_\omega^2$) and the $\eta$ transition 
form factor (for $q^2=0$ and $q^2=M_\phi^2$) are shown in 
Table~\ref{tab:lambdEx},
\begin{table}[t]
\renewcommand{\arraystretch}{1.3}
\begin{center}
\begin{tabular}{lll}
\hline
\multicolumn{3}{c}{$b_{\pi^0}(q^2)$ $[$GeV$^{-2}]$}\\
\hline
$b_{\pi^0}(q^2=0)$  & 1.79 $\pm$ 0.14 & CELLO \cite{Behrend:1990sr}\\
$b_{\pi^0}(q^2=0.613 $GeV$^2)$  & 2.4 $\pm$ 0.2& Lepton-G \cite{Dzhelyadin:1980tj}\\
$b_{\pi^0}(q^2=0.613 $GeV$^2)$  & 2.24 $\pm$ 0.06 $\pm$ 0.02& NA60 \cite{Arnaldi:2009wb}\\
$b_{\pi^0}(q^2=0.613 $GeV$^2)$  & 2.241 $\pm$ 0.025 $\pm$ 0.028& NA60 \cite{Usai:2011zza}\\
\hline
\multicolumn{3}{c}{$b_{\eta}(q^2)$ $[$GeV$^{-2}]$}\\
\hline
$b_{\eta}(q^2=0)$  & 1.9 $\pm$ 0.4& Lepton-G \cite{Dzhelyadin:1980kh}\\
$b_{\eta}(q^2=0)$  & 1.42 $\pm$ 0.21 & CELLO \cite{Behrend:1990sr}\\
$b_{\eta}(q^2=0)$  & 1.95 $\pm$ 0.17 $\pm$ 0.05& NA60 \cite{Arnaldi:2009wb}\\
$b_{\eta}(q^2=0)$  & 1.950 $\pm$ 0.059 $\pm$ 0.042 & NA60 \cite{Usai:2011zza}\\
$b_{\eta}(q^2=0)$  & 1.92 $\pm$ 0.35 $\pm$ 0.13& CB/TAPS \cite{Berghauser:2011zz}\\
$b_{\eta}(q^2=1.040 $GeV$^2)$  & 3.8 $\pm$ 1.8& SND \cite{Achasov:2000ne}\\
\hline
\end{tabular}
\end{center} \vspace{-3mm}
\caption{Summary of the available experimental data on $b_{\eta}(q^2)$ and $b_{\pi^0}(q^2)$. \label{tab:lambdEx}}
\end{table}
and their $q^2$ dependence compared to the VMD model in Fig.~\ref{fig:slopes}.
Single off-shell form factors can be studied in:
\begin{enumerate} 
\item pseudoscalar meson decays $P\to l^+l^-\gamma$   
 ($4m_l^2<q^2<m^2_P$);
\item $e^+e^-\to P\gamma$   ($q^2>m^2_P$), where the cross section is given by
\begin{equation}
\sigma(e^+e^-\to P\gamma)=4\pi\alpha_{\rm QED}\Gamma_{\gamma\gamma}
\left(\frac{s-m_P^2}{sm_P}\right)^3|F_P(q^2=s,0)|^2 \label{eq:ee-Pg} ~;
 \end{equation}
\item two-photon production: e.g., in $e^- e^\pm$ interactions
or the Primakoff process ($q^2<0$)
\begin{equation}
\sigma_{\gamma^*\gamma^*\to P}=\frac{16\pi^2}{m_P^3}\Gamma_{\gamma\gamma}
|F(q_1^2,q_2^2)|^2
\sqrt{(q_1\cdot q_2)^2-q_1^2q_2^2}\ \delta\left((q_1+ q_2)^2-m_P^2\right) ~.
\end{equation}
\end{enumerate}

\subsection{Dalitz decays}

\begin{figure}[t]
\centering
\includegraphics[width=0.3\textwidth]{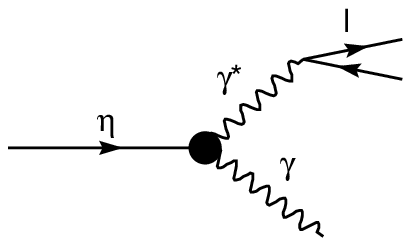} \hspace{1cm}
\includegraphics[width=0.3\textwidth]{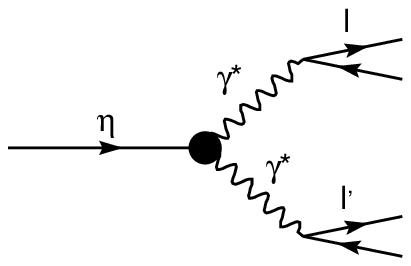}
\caption{\label{fig:dalitz} Single and double Dalitz decays}
\end{figure}

The  $q_{1,2}^2$ for the  conversion (Dalitz)
decays, see Fig.~\ref{fig:dalitz}, is equal to the invariant mass squared 
of the lepton-antilepton pair, and $m_P^2\ge q_{1,2}^2\ge 4m_\ell^2$  
(time-like virtual  photons).  
The amplitude of  the single conversion decay of  a pseudoscalar meson
$P$ is given by
\begin{equation}
{\mathcal A}(P\to\ell^+\ell^-\gamma)= ie{\cal F}_P(q_1^2,0)
\epsilon_{\mu\nu\sigma\tau}\varepsilon_2^\mu q_2^\nu\varepsilon_1^{\sigma}\ 
  \frac{1}{q_1^2}\ 
[\bar{u}\gamma^\tau u] ~,
\end{equation}
where  $1/q_1^2$ is the  photon propagator  and the  last term  is the
leptonic current.  

\begin{figure}[t]
\centering
\includegraphics[width=0.6\textwidth]{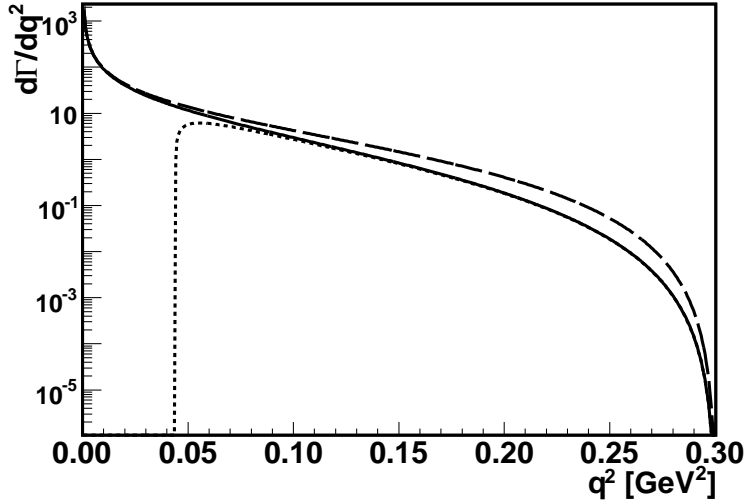}
\caption{\label{fig:daldec}   $d\Gamma/dq^2$   distributions  for the
  single Dalitz  decays of  the $\eta$ meson. The solid  line corresponds to
  $\eta\to  e^+e^-\gamma$  with $F_\eta(q^2,0)=1$;  the dotted  line shows  $\eta\to
  \mu^+\mu^-\gamma$   with  $F_\eta(q^2)=1$;  finally, the dashed   line  is   $\eta\to
  e^+e^-\gamma$  with  the  VMD  form  factor, see Sect.~\ref{sec:VMD}.
}
\end{figure}

Experimentally, the  form factor can be extracted   from the $q^2=q_1^2$
distribution given by
\begin{equation}
\frac{d\Gamma(P\to  \ell^+\ell^- \gamma)}{dq^2 \Gamma_{\gamma\gamma}} =
\frac{2\alpha}{3\pi}\frac{1}{q^2}\sqrt{1-\frac{4m^2_\ell}{q^2}}
\left(1+\frac{2m^2_\ell}{q^2}\right)
\left(1-\frac{q^2}{M^2_P}\right)^3|F_P(q^2,0)|^2.
\label{eqn:dgdq2}
\end{equation}
The distributions  for $\eta$ Dalitz  decays are presented
in  Fig.~\ref{fig:daldec}.   The  distributions for  the  $e^+e^-\gamma$
final states are peaked at $4m_\ell^2$ due to the $1/q^2$ QED term.  
The  form factor  can be   obtained  by dividing  out  the QED  dependence.
To extract the form factor slope, a  dependence on the  $q^2$ variable  
is often fitted
with a single-pole formula, which at low energies $q^2\ll\Lambda^2$ 
can be approximated successively as follows:
\begin{equation}
F(q^2,0)= 
\frac{\Lambda^2}{\Lambda^2-q^2-i\Gamma\Lambda} 
\approx \frac{\Lambda^2}{\Lambda^2-{q^2}} 
\approx 1+\frac{q^2}{\Lambda^2}.
\label{eqn:pole}
\end{equation}
The form factor slope $b_P(0)$ is therefore related to $\Lambda$:
\begin{equation}
b_P(0)\equiv\left.\frac{d\ln F(q^2,0)}{dq^2}\right|_{q^2=0}=\frac{1}{\Lambda^2} ~.
\label{eqn:slope}
\end{equation}

\subsection{Radiative and hadronic anomalous processes}

Finally, there is one more twist: the $\rho$ meson is rather broad and
couples  to two pions.  Therefore processes that involve  two pions
instead of a  dilepton are also intimately connected  to the previous
reactions.    This   includes,   e.g.,    $\eta   \to    \pi^+   \pi^-
\gamma^{(*)}$. There are indications for deviations from vector-meson
dominance~\cite{Adlarson:2011xb,stollenwerk,Zdebik}. 
(Of  course,   this  logic  can  be
extended  to reactions  with three  pions  instead of  an $\omega$  or
$\phi$. In these cases, however, there are a lot of different channels 
interfering with each other.)
--- A list of possible reactions thus related to each other
is provided in Appendix~1.A.

\subsection{A naive Vector Meson Dominance model picture}\label{sec:VMD}

For some processes, model-independent analyses using dispersion theory are possible;
for recent developments, see e.g.\ Ref.~\cite{Schneider:2012ez,Stollenwerk:WIP}.
Symmetry constraints may be invoked using  properly-chosen matching conditions, say,
to chiral perturbation theory~\cite{stollenwerk}.
However here, for illustration, only predictions of
the most  naive VMD  model (photon couplings {\it solely} through vector mesons)
 will be presented  in some  detail. The transition form
factor of a pseudoscalar meson $P(\pi^0,\eta,\eta')$ is given within the
model by~\cite{Feynman:1989aa,Bramon:1981sw,Landsberg:1986fd}:
\begin{equation}
   F_P^{\rm VMD}(q_1^2,q_2^2)= \frac{1}{N}\sum_V\sum_{V'} \frac{g_{PVV'}}{g_{V\gamma}g_{V'\gamma}} 
\frac{D_V(0)}{D_V(q_1^2)}\frac{D_{V'}(0)}{D_{V'}(q_2^2)}.
\label{eqn:FVMD}
\end{equation}
The  sum extends over  the neutral  vector mesons  
($Q=S=0$): $\rho^0$,
$\omega$, $\phi$, \ldots; $g_{PVV'}$ and  $g_{V\gamma}$ 
are their flavor SU(3) couplings to 
the pseudoscalar meson $P$ and to the  photon, respectively.
The functions  $D_V(q^2)$ are vector meson propagators, where for 
illustration we use the simplest expression defined 
in the whole (real) $q^2$ range: 
\begin{equation}
D_V(q^2)=M_V^2-q^2-i\Gamma_VM_V
\end{equation}
with constant vector meson widths $\Gamma_V$. 
$D_V(0)$ is approximately $M_V^2$ as long as the width is considered small.

\begin{figure}[t]
\centering
\includegraphics[width=0.95\textwidth]{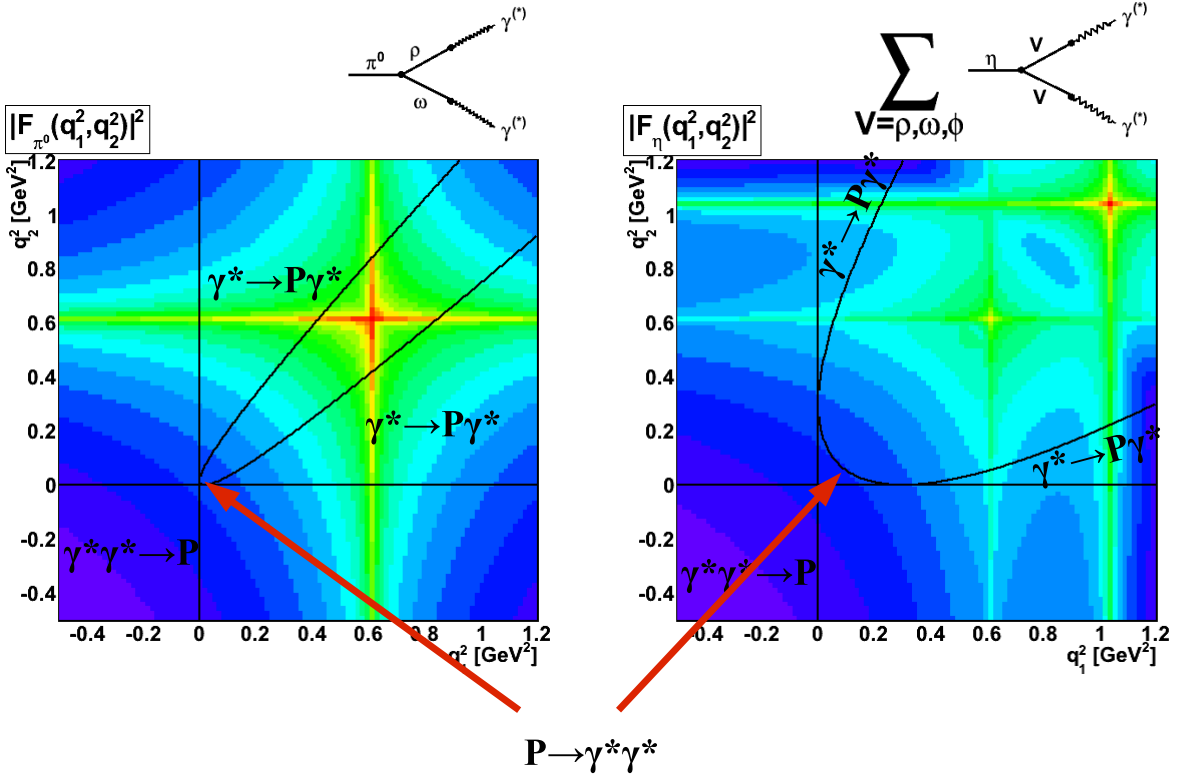}
\caption{\label{fig:ff2dpl}The $\pi^0$ and $\eta$ meson form factor 
squared in naive VMD.}
\end{figure}

In the following, we keep
only the three lightest vector mesons  in the sum and take the values
of the  couplings $g_{\eta VV}$  and \gvga as expected from the
quark model; moreover, the $\eta$ has been assumed to be a pure flavor-octet 
state for simplicity:
\begin{equation}
 \frac{1}{2g_{\rho\gamma}}: \frac{1}{2g_{\omega\gamma}}: 
\frac{1}{2g_{\phi\gamma}} = 
1: \frac{1}{3} : -\frac{\sqrt{2}}{3} ~, \quad
{g_{\eta\rho\rho}}:{g_{\eta\omega\omega}}:{g_{\eta\phi\phi}} = 
1 : 1 : -2 ~.
\end{equation}
These factors are obtained in the following way: using the meson multiplets
\begin{eqnarray}
  P = \left(
    \begin{array}{ccc}
      \pi^0+\frac{1}{\sqrt{3}}\eta & \sqrt{2}\pi^+ & \sqrt{2}K^+ \\
      \sqrt{2}\pi^- & -\pi^0+\frac{1}{\sqrt{3}}\eta & \sqrt{2}K^0 \\
      \sqrt{2}K^- & \sqrt{2}\bar{K}^0 & -\frac{2}{\sqrt{3}}\eta
    \end{array}
  \right)
, \quad 
  V = \left(
    \begin{array}{ccc}
      \rho^0 + \omega & \sqrt{2}\rho^+ & \sqrt{2}K^{*+} \\
      \sqrt{2}\rho^- & -\rho^0 + \omega & \sqrt{2}K^{*0} \\
      \sqrt{2}K^{*-} & \sqrt{2}\bar{K}^{*0} & \sqrt{2}\phi
    \end{array}
  \right),
  \label{eq:p+vmult}
\end{eqnarray}
and the quark-charge matrix 
\begin{eqnarray}
  Q = \left(
    \begin{array}{ccc}
      \frac{2}{3} & 0 & 0 \\
      0 & -\frac{1}{3} & 0 \\
      0 & 0 & -\frac{1}{3}
    \end{array}
  \right),
  \label{eq:quarkcharge}  
\end{eqnarray}
one obtains the photon coupling ratios from the flavor trace 
${\rm tr}\{V \, Q\}$ and the $PVV$ couplings from
${\rm tr}\{(V_1 \, V_2 + V_2 \, V_1) \, P\}$. 
The normalization factor $N$ ensures that $F_P^{\rm VMD}(0,0)=1$.
For $\pi^0$ and $\eta$, due to isospin conservation and the OZI rule
one finds the following expressions: 
\begin{align}
   F_{\pi^0}^{\rm VMD}(q_1^2,q_2^2)&= \frac{1}{2}\left\{ 
\frac{D_\rho(0)}{D_\rho(q_1^2)}
\frac{D_{\omega}(0)}{D_{\omega}(q_2^2)}+\frac{D_{\omega}(0)}{D_{\omega}(q_1^2)}
\frac{D_\rho(0)}{D_\rho(q_2^2)}
\right\} ~,  \label{eqn:FVMDpi+et}\\
   F_{\eta}^{\rm VMD}(q_1^2,q_2^2)&= \frac{1}{N}\left\{ 
\frac{g_{\eta\rho\rho}}{g^2_{\rho\gamma}}
\frac{D_\rho(0)}{D_\rho(q_1^2)}
\frac{D_\rho(0)}{D_\rho(q_2^2)}\right.+
\frac{g_{\eta\omega\omega}}{g^2_{\omega\gamma}}
\frac{D_{\omega}(0)}{D_{\omega}(q_1^2)}
\frac{D_{\omega}(0)}{D_{\omega}(q_2^2)}
+
\left.\frac{g_{\eta\phi\phi}}{g^2_{\phi\gamma}}
\frac{D_{\phi}(0)}{D_{\phi}(q_1^2)}
\frac{D_{\phi}(0)}{D_{\phi}(q_2^2)}
\right\} ~.
\notag
\end{align}
In Fig.~\ref{fig:ff2dpl} the  resulting squares of the absolute values 
of the form factors 
are drawn for $\pi^0$ and  $\eta$. There are three experimentally 
accessible
regions of the form factors.  The  boundaries are defined by the parabola and
the axes of the plots. The region inside the parabola and inside the 
second and fourth 
quarters of the diagrams are not accessible experimentally.
The $P\to\gamma^{(*)}\gamma^{(*)}$ decays will probe the  central region 
of the plot defined by the conditions: $q_{1,2}^2 \ge 0$ and 
$\sqrt{q_1^2}+\sqrt{q_2^2} \le m_P$.
Processes $e^+e^-\to P\gamma^{(*)}$
correspond to: $\sqrt{q_1^2}>m_P$, $0 \le \sqrt{q_2^2} \le \sqrt{q_1^2}-m_P$. 
Finally, $\gamma^{(*)}\gamma^{(*)}\to P$
processes (from $e^+ e^-$ collisions) allow to study the 
whole $q_1^2,q_2^2\leq 0$ region. 

In particular, for a single on-shell photon the form factor reduces to
the formula from Ref.~\cite{Landsberg:1986fd}:
\begin{equation}
   F_P^{\rm VMD}(q^2,0)= \frac{1}{N}\sum_V \frac{g_{PV\gamma}}{2g_{V\gamma}} 
\frac{D_V(0)}{D_V(q^2)}\approx  \frac{1}{N}\sum_V \frac{g_{PV\gamma}}{2g_{V\gamma}} 
\frac{M_V^2}{D_V(q^2)} ~,
\label{eqn:FVMDo}
\end{equation}
where 
\begin{equation}
 g_{P V \gamma}=\sum_{V'}\frac{g_{PVV'}}{g_{V'\gamma}} ~.
\end{equation}

\subsection*{Appendix}

\setcounter{subsection}{0}
\renewcommand{\thesubsection}{\thesection.\Alph{subsection}}
\subsection{List of processes}
\renewcommand{\thesubsection}{\thesection.\arabic{subsection}}

Here, we compile a list of reactions that are relevant and 
linked to each other 
in the context of transition form factors. 
The focus is on reactions that are possible to study in practice.
For example, scattering reactions with initial-state neutral pions 
(like $\pi^0 \gamma \to $ something) are not considered
because there are no $\pi^0$ beams available. There is a charged-pion beam, 
though (COMPASS at CERN), and ``collisions'' with 
photons can be achieved using the Primakoff effect. Likewise, reactions with 
initial-state muons are not listed, though
this is one of the future dreams of high-energy physicists. 
Reactions that involve a broad $\rho$ meson are listed in the following 
as $\pi^+ \pi^-$. In principle, reactions like  $e^+ e^- \to \pi^+ \pi^-$ 
and $e^+ e^- \to \pi^0 \omega$ 
are related to $\tau^- \to \nu_\tau \, \pi^- \pi^0$ and
 $\tau^- \to \nu_\tau  \, \pi^- \omega$ through conservation of vector 
current and isospin symmetry~\cite{Tsai:1971vv}. 
Decays of the $\tau$ are not listed below. 
\begin{enumerate}
\item Processes with one external (pseudoscalar) hadron: $\pi^0$

\begin{tabular}{ll}
\begin{minipage}{0.47\linewidth}
  \begin{enumerate}
  \item $\pi^0 \to 2 \gamma$
  \item $\pi^0 \to \gamma \, e^+ e^-$
  \item $\pi^0 \to e^+ e^- \, e^+ e^-$
  \item $\pi^0 \to e^+ e^-$
  \end{enumerate}
\end{minipage}
&
\begin{minipage}{0.47\linewidth}
  \begin{enumerate}
  \addtocounter{enumii}{4}
  \item $e^+ e^- \to \pi^0 \, \gamma$
  \item $e^+ e^- \to \pi^0 \, e^+ e^-$
  \item $e^+ e^- \to \pi^0 \, \mu^+ \mu^-$ 
  \item[]
  \end{enumerate}
\end{minipage}
\end{tabular}

\item Processes with one external (pseudoscalar) hadron: $\eta$

\begin{tabular}{ll}
\begin{minipage}{0.47\linewidth}
  \begin{enumerate}
  \item $\eta \to 2 \gamma$
  \item $\eta \to \gamma \, e^+ e^-$
  \item $\eta \to \gamma \, \mu^+ \mu^-$
  \item $\eta \to e^+ e^- \, e^+ e^-$
  \item $\eta \to e^+ e^- \, \mu^+ \mu^-$
  \item $\eta \to \mu^+ \mu^- \, \mu^+ \mu^-$
  \end{enumerate}
\end{minipage}
&
\begin{minipage}{0.47\linewidth}
  \begin{enumerate}
  \addtocounter{enumii}{6}
  \item $\eta \to e^+ e^-$
  \item $\eta \to \mu^+ \mu^-$
  \item $e^+ e^- \to \eta \, \gamma$
  \item $e^+ e^- \to \eta \, e^+ e^-$
  \item $e^+ e^- \to \eta \, \mu^+ \mu^-$
  \item[]
  \end{enumerate}
\end{minipage}
\end{tabular}

\item Processes with one external (pseudoscalar) hadron: $\eta^{\prime}$

\begin{tabular}{ll}
\begin{minipage}{0.47\linewidth}
  \begin{enumerate}
  \item $\eta^\prime \to 2 \gamma$
  \item $\eta^\prime \to \gamma \, e^+ e^-$
  \item $\eta^\prime \to \gamma \, \mu^+ \mu^-$
  \item $\eta^\prime \to e^+ e^- \, e^+ e^-$
  \item $\eta^\prime \to e^+ e^- \, \mu^+ \mu^-$
  \item $\eta^\prime \to \mu^+ \mu^- \, \mu^+ \mu^-$
  \end{enumerate}
\end{minipage}
&
\begin{minipage}{0.47\linewidth}
  \begin{enumerate}
  \addtocounter{enumii}{6}
  \item $\eta^\prime \to e^+ e^-$
  \item $\eta^\prime \to \mu^+ \mu^-$
  \item $e^+ e^- \to \eta^\prime  \, \gamma$
  \item $e^+ e^- \to \eta^\prime  \, e^+ e^-$
  \item $e^+ e^- \to \eta^\prime  \, \mu^+ \mu^-$
  \item[]
  \end{enumerate}
\end{minipage}
\end{tabular}

\item Processes with two external (narrow) hadrons: $\pi^0$ and a vector meson

\begin{tabular}{ll}
\begin{minipage}{0.47\linewidth}
  \begin{enumerate}
  \item $\omega \to \pi^0 \, \gamma$
  \item $\omega \to \pi^0 \, e^+ e^-$
  \item $\omega \to \pi^0 \, \mu^+ \mu^-$
  \item $e^+ e^- \to \omega \, \pi^0$
  \end{enumerate}
\end{minipage}
&
\begin{minipage}{0.47\linewidth}
  \begin{enumerate}
  \addtocounter{enumii}{4}
  \item $\phi \to \pi^0 \, \gamma$
  \item $\phi \to \pi^0 \, e^+ e^-$
  \item $\phi \to \pi^0 \, \mu^+ \mu^-$
  \item $e^+ e^- \to \phi \, \pi^0$
  \end{enumerate}
\end{minipage}
\end{tabular}

\item Processes with two external (narrow) hadrons: $\eta$ and a vector meson
\nopagebreak

\begin{tabular}{ll}
\begin{minipage}{0.47\linewidth}
  \begin{enumerate}
  \item $\omega \to \eta \, \gamma$
  \item $\omega \to \eta \, e^+ e^-$
  \item $\omega \to \eta \, \mu^+ \mu^-$
  \item $e^+ e^- \to \omega \, \eta$
  \end{enumerate}
\end{minipage}
&
\begin{minipage}{0.47\linewidth}
  \begin{enumerate}
  \addtocounter{enumii}{4}
  \item $\phi \to \eta \, \gamma$
  \item $\phi \to \eta \, e^+ e^-$
  \item $\phi \to \eta \, \mu^+ \mu^-$
  \item $e^+ e^- \to \phi \, \eta$
  \end{enumerate}
\end{minipage}
\end{tabular}

\item Processes with two external (narrow) hadrons: $\eta^\prime $ 
and a vector meson

\begin{tabular}{ll}
\begin{minipage}{0.47\linewidth}
  \begin{enumerate}
  \item $\eta^\prime  \to \omega \, \gamma$
  \item $\eta^\prime  \to \omega  \, e^+ e^-$
  \item $e^+ e^- \to \omega \, \eta^\prime $
  \end{enumerate}
\end{minipage}
&
\begin{minipage}{0.47\linewidth}
  \begin{enumerate}
  \addtocounter{enumii}{3}
  \item $\phi \to \eta^\prime  \, \gamma$
  \item $\phi \to \eta^\prime \, e^+ e^-$
  \item $e^+ e^- \to \phi \, \eta^\prime$
  \end{enumerate}
\end{minipage}
\end{tabular}

\item Processes with three external (pseudoscalar) hadrons: 
$\pi^0 \, \pi^+ \, \pi^-$

\begin{tabular}{ll}
\begin{minipage}{0.47\linewidth}
  \begin{enumerate}
  \item $\pi^\pm \, \gamma \to \pi^0 \, \pi^\pm$
  \end{enumerate}
\end{minipage}
&
\begin{minipage}{0.47\linewidth}
  \begin{enumerate}
  \addtocounter{enumii}{1}
  \item $ e^+ e^- \to \pi^0 \, \pi^+ \, \pi^-$
  \end{enumerate}
\end{minipage}
\end{tabular}

\item Processes with three external (pseudoscalar) hadrons: 
$\eta \, \pi^+ \, \pi^-$

\begin{tabular}{ll}
\begin{minipage}{0.47\linewidth}
  \begin{enumerate}
  \item $\eta \to \pi^+ \pi^- \, \gamma$ 
  \item $\eta \to \pi^+ \pi^- \, e^+ e^-$
  \item $\eta \to \pi^+ \pi^- \, \mu^+ \mu^-$
  \end{enumerate}
\end{minipage}
&
\begin{minipage}{0.47\linewidth}
  \begin{enumerate}
  \addtocounter{enumii}{3}
  \item $\pi^\pm \, \gamma \to \eta \, \pi^\pm$
  \item $ e^+ e^- \to \eta \, \pi^+ \, \pi^-$
  \item[]
  \end{enumerate}
\end{minipage}
\end{tabular}

\item Processes with three external (pseudoscalar) hadrons: 
$\eta^\prime \, \pi^+ \, \pi^-$

\begin{tabular}{ll}
\begin{minipage}{0.47\linewidth}
  \begin{enumerate}
  \item $\eta^\prime \to \pi^+ \pi^- \, \gamma$ 
  \item $\eta^\prime \to \pi^+ \pi^- \, e^+ e^-$
  \item $\eta^\prime \to \pi^+ \pi^- \, \mu^+ \mu^-$
  \end{enumerate}
\end{minipage}
&
\begin{minipage}{0.47\linewidth}
  \begin{enumerate}
  \addtocounter{enumii}{3}
  \item $\pi^\pm \, \gamma \to \eta^\prime \, \pi^\pm$
  \item $ e^+ e^- \to \eta^\prime \, \pi^+ \, \pi^-$
  \item[]
  \end{enumerate}
\end{minipage}
\end{tabular}

\end{enumerate}
Note that the coupling of three neutral pseudoscalars to a photon 
breaks charge-conjugation invariance. 
In principle, one could add processes with more external 
hadrons to the list, e.g., 
$\eta^\prime \to \pi^+ \pi^- \, \pi^+ \pi^-$ and 
$\eta^\prime \to \pi^+ \pi^0 \, \pi^- \pi^0$, which could proceed via two
virtual $\rho$ mesons. For a theoretical discussion of these decays,
see Ref.~\cite{guokubiswirzba}. The list above is restricted to four or less external 
states, but counting a virtual photon as one state.

\renewcommand\refname

\newpage

\section{Short summaries of the talks}

\subsection[Muon $g-2$ from $\tau$ and $e^+ e^-$ data]{Muon \boldmath{$g-2$} from \boldmath{$\tau$} and \boldmath{$e^+e^-$} data}
\addtocontents{toc}{\hspace{2cm}{\sl F.~Jegerlehner}\par}

\vspace{5mm}

\noindent
F.~Jegerlehner

\vspace{5mm}

\noindent
Humboldt University Berlin and DESY Zeuthen

\vspace{5mm} 
The leading hadronic contribution to the very precisely measured muon
$g-2$~\cite{BNL06} is obtained most reliably by using a dispersion relation
together with $\epm \to$ hadrons data. As a drawback, the precision of
the theoretical prediction of the muon anomaly $a_\mu$ is dominated by
experimental errors. The dominant contribution is coming from the
$\pi^+\pi^-$ channel at low energies. More data help to improve the
precision of the prediction.  The hadronic final states in the
reactions $e^+e^- \to \pi^+\pi^-$ and $\tau^- \to \nu_\tau \pi^- \pi^0$
are related by isospin symmetry, up to isospin breaking (IB) effects
due to $m_d \neq m_u$ and differently charged particles
involved. After an isospin rotation and IB-corrections one thus can
use the high quality $\tau$--decay spectra to enhance the data set
needed for evaluating $\amuh$~\cite{ADH98}. As more data and data of
better quality got available, a serious clash showed up between $\tau$-
and $\epm$-data~\cite{DEHZ03}. Recently, in Ref.~\cite{JS11}, the
$\tau$ vs. $\epm$ puzzle has been resolved. The discrepancy turned out
to be due to so far unaccounted self-energy corrections in
$\rho^0-\gamma$ mixing. It contributes to $e^+e^- \to \pi^+\pi^-$, but
is absent in $\tau$-decays. After correcting the $\tau$-data for the
missing $\rho-\gamma$ mixing, besides the other known IB-corrections,
the I=1 $\pi\pi$ results are fully compatible. $\tau$-data thus
confirm $\epm$-based $\amuh$ evaluations. For the
leading-order vacuum polarization contribution, based on all $\epm$
data (CMD2, SND, KLOE and BaBar), we obtain
$a_\mu^{\mathrm{had,LO}}[e]=690.75(4.72)\power{-10}$ ($\epm$ based),
while including $\tau$ data (ALEPH, OPAL, CLEO and Belle) we find
$a_\mu^{\mathrm{had,LO}}[e,\tau]=690.96(4.65)\power{-10}$
($\epm$+$\tau$ based). This backs the $\sim$3$\,\sigma$ deviation
between $\amu^{\mathrm{experiment}}$ and $\amu^{\mathrm{theory}}$. For
the $\tau$ di-pion branching fraction we find
$B^{\mathrm{CVC}}_{\pi\pi^0}=25.20\pm 0.17\pm 0.28$ from $\epm$+CVC,
while $B_{\pi\pi^0}=25.34\pm 0.06\pm 0.08$ is evaluated directly from
the $\tau$ spectra. Applying our correction to the most recent
evaluation~\cite{BABAR12} we have
\begin{center}
\renewcommand{\arraystretch}{1.3}
\begin{tabular}{ccccc}
\hline
$a_\mu^{2\pi,\mathrm{LO}}[E<1.8~\gv]$ & $\amu^{\mathrm{the}}$ &
$\amu^{\mathrm{exp}}-\amu^{\mathrm{the}}$ & & \\
\hline
\mbo{(514.1 \pm 3.8)} &\mbo{11659186.5(5.4)} &\mbo{(22.4 \pm 8.3)}&\mbo{2.7\,\sigma} &[BaBar] \\
\mbo{(507.8 \pm 3.2)} &\mbo{11659180.2(5.0)} &\mbo{(28.7 \pm 8.0)}&\mbo{3.6\,\sigma} &[$ee$ all] \\
\mbo{(508.7 \pm 2.5)} &\mbo{11659181.1(4.6)} &\mbo{(27.8 \pm 7.8)}&\mbo{3.6\,\sigma} &[$ee + \tau$] \\
\hline
\end{tabular}
\end{center}
\noindent
(in units $10^{-10}$) where the last entry is our estimate combining
$ee$ and $\tau$ results (see also~\cite{HLS12}).

\newpage

\subsection{A lattice study of the leading order hadronic contribution to 
the muon anomalous magnetic moment}
\addtocontents{toc}{\hspace{2cm}{\sl K.~Jansen}\par}

\vspace{5mm}

\noindent
X.~Feng, G.~Hotzel, \underline{K.~Jansen}, M.~Petschlies, D.~B.~Renner

\vspace{5mm}
 
\noindent

\vspace{5mm} 
We present a calculation 
of the leading order hadronic contribution to the muon anomalous magnetic moment
$a_\mu^{\rm had}$ from lattice QCD computations based on first principles. 
We report about a first benchmark calculation with mass-degenerate 
up and down quarks using the ETMC configurations~\cite{Baron:2009wt}.
Using maximally twisted mass fermions as our formulation 
of lattice QCD guarantees that  
physical observables are automatically accurate to $O(a^2)$ in the 
lattice spacing,
allowing therefore a well 
controlled continuum limit. 
By employing an improved definition of the lattice observable
for $a_\mu^{\rm had}$~\cite{Feng:2011zk,Feng:lattice} we can achieve an accuracy
of $a_\mu^{\rm had}$ that is almost matching the experimental precision.
Using this improved method allows us to also accurately 
compute the hadronic contribution to 
$\Delta\alpha_{\rm QED}$, the Adler function and muonic atoms, see
\cite{Feng:lattice}. 
Since ETMC has by now also calculations with a 
dynamical strange and charm quark\cite{Baron:2010bv}, 
we have also performed a first ever four-flavor calculation 
of $a_\mu^{\rm had}$ finding consistent values with phenomenological
standard model calculations.

\newpage

\subsection{Confronting the scan, ISR and tau dipion spectra 
within a global model}
\addtocontents{toc}{\hspace{2cm}{\sl M.~Benayoun}\par}

\vspace{5mm}

\noindent
M.~Benayoun

\vspace{5mm}

\noindent
LPNHE des Universit\'es Paris VI et Paris VII

\vspace{5mm} 
The Hidden Local Symmetry (HLS) Model~\cite{HLSRef} has given rise to successful
phenomenological studies once a mechanism accounting for the breaking of the
SU(3) and SU(2) symmetries has been performed. An additional mechanism 
has been implemented which performs the mixing of the vector mesons
$\rho^0$, $\omega$ and $\phi$; this vector meson mixing plays a 
fundamental role in generating
the coupling of the $\omega$ and $\phi$ mesons to a pion pair.
Within this framework (which can be found in detail in Ref.\cite{ExtMod3}), 
it is
possible to treat consistently and $simultaneously$ the annihilation channels
$e^+e^-\to \pi^+\pi^-/K^+K^-/K^0 \overline{K}^0/\pi^0 \gamma 
/\eta \gamma /\pi^+\pi^-\pi^0$, the $\tau^\pm \to \pi^\pm \pi^0 \nu_\tau$ 
decay spectrum and some additional decay partial widths. Using  almost all the
relevant existing data sets collected in scan experiments 
and the ALEPH, CLEO and
BELLE $\tau$ spectra, one has reached a surprisingly good 
accuracy as illustrated in~\cite{ExtMod1,ExtMod3}.

We have presented an update of the analysis reported in Ref.~\cite{ExtMod3}
by examining the behavior of the $e^+e^-\to \pi^+\pi^-$ spectra collected
at the VEPP2M $e^+e^-$ collider by the scan procedure and those collected 
by KLOE (referred to as KLOE08 and KLOE10) and also by BaBar 
using the Initial State Radiation (ISR) method. 
The tool for this comparison is a global fit including the three $\tau$ spectra
already mentioned together with all data referred to above (excluding  
$e^+e^-\to \pi^+\pi^-$). However, this is not sufficiently precise and using some 
more information (essentially the $\rho^0 \to e^+e^-$ partial width, marginally
present in the other channels considered, and the $\omega/\phi \to \pi^+\pi^-$ 
partial widths), it has been possible to show that the $e^+e^-\to \pi^+\pi^-$
cross sections from CMD-2, SND and KLOE10 are perfectly consistent 
with each other and
with all the other (non--$\pi^+\pi^-$) data considered. 

We have shown that this (reference) merged data set (CMD-2, SND and KLOE10)
introduced within the global fit code leads to a value for the muon
$g-2$ which is at about $4.5 \sigma$ from the BNL measurement~\cite{BNL2}. 
The returned global fit quality is above the 90\% level. On the other hand,
using some weighting, the KLOE08 and BaBar data sets have also been considered
within the global fit; this has confirmed the above mentioned result 
for the muon $g-2$.

\newpage

\subsection{Light Meson Decays with WASA-at-COSY}
\addtocontents{toc}{\hspace{2cm}{\sl H.~Bhatt}\par}

\vspace{5mm}

\noindent \underline{H.~Bhatt}$^{a,b}$ and S.~Schadmand$^{b}$ (for the WASA-at-COSY Collaboration)

\vspace{5mm}

\noindent
$^{a}$IIT Bombay and $^{b}$Forschungszentrum J\"ulich

\vspace{5mm}

The meson decay physics program with the WASA detector at the 
COSY accelerator pursues symmetry
breaking processes by studying rare decays of the light unflavored mesons 
as well as the
determination of electromagnetic transition form factors. 
The experimental approach uses the WASA
facility which is a $4\pi$ detector system, designed to study 
the hadronic production and the
decays of light mesons with the decay products $\gamma,\pi, e^\pm$. 
The unique high density pellet target combined with high intensity beams 
of the Cooler Synchrotron COSY, J\"ulich, provide
luminosities which allow studies of rare processes~\cite{ref:wasa}.

The three-pion decays of $\eta$ and $\omega$ mesons hold important 
information for chiral
perturbation theory and the promise to extract limits on the quark mass 
differences. For the $\eta$
decay into three neutral pions, the Dalitz plot of the three pions 
was studied and the quadratic
slope parameter $\alpha$ was determined to be $-0.027 \pm 0.008(stat) \pm 0.005(syst)$. The result
is consistent with previous measurements and further corroborates the importance of pion-pion final
state interactions~\cite{ref:Vlasov}. Using our extended data set 
on $\eta$ decays as well as
recently acquired data on $\omega$ decays, we continue the Dalitz plot analyses for the case of
$\pi^+\pi^-\pi^0$ in the final state~\cite{ref:MESON,ref:primenet}.

Decays of the $\pi^0$ meson allow to search for gauge bosons mediating 
dark forces in the MeV
range. The decay $\pi^0\to ee\gamma$ is sensitive to 
a ``dark photon'' that decays into an $e^+e^-$
pair. WASA-at-COSY has collected a 500k data sample to constrain 
the parameters of this
hypothetical gauge boson~\cite{ref:Gullström}. The rare decay 
$\pi^0\to ee$ could probe physics
beyond the Standard Model. Deviations between experiment 
and the Standard Model prediction would be
explained by a dark gauge boson, which might also account for 
the enhanced $e^+e^-$ annihilation
line from the galactic center. The upcoming high statistics 
run with WASA-at-COSY could confirm the
present experimental result.

We have performed an exclusive measurement of the decay 
$\eta\to\pi^+\pi^-\gamma$. At the chiral
limit of zero momentum and massless quarks this decay 
is determined by the box anomaly term of the
Wess--Zumino--Witten Lagrangian. The measured pion angular 
distribution is consistent with a relative
p-wave of the two-pion system, whereas the measured photon energy spectrum 
was found at variance
with the simplest gauge invariant matrix element. 
A parameterization of the data can be achieved by
the additional inclusion of the empirical pion vector form factor 
multiplied by a first-order
polynomial in the squared invariant mass of the 
$\pi\pi$ system~\cite{ref:Redmer}.

We report on our efforts to determine transition form factors 
using the decays $\eta\to\gamma^{(*)}ee$ and $\omega\to\pi ee$. 
For the status and preliminary results on transition form factors, we
present preliminary results for the $\eta$ transition form factor 
from the decays $\eta\to\gamma ee$. The results are based on two 
independent analyses using two different production
mechanisms~\cite{ref:Hodana,ref:Bhatt}. We point out further steps 
towards our final result,
expected at the end of 2012. The decay $\eta\to eeee$ carries 
information about the double
transition form factor and is being studied with the first goal 
of a branching ratio~\cite{ref:Yurev,ref:primenet}. $\omega$ decays 
are being analyzed using a four-week $pd\to {}^3$He$\,\omega$ experiment 
and a  one-week $p+p$ pilot run. As a first step towards the $\omega\pi$
transition form factor, we study the decay with a real photon, 
$\omega\to\pi\gamma$. Promising
first results show the $\omega$ peak in a missing mass analysis, 
inclusively and in the
presence of the respective decay products~\cite{ref:MESON,ref:primenet}.

In summary, we have accumulated $3\times 10^7$ $\eta$ mesons 
produced in $pd\to {}^3$He$\,\eta$ and
5--10$\times 10^8$ $\eta$ decays in $pp\to pp\eta$. The experimental goal is 
to study very rare meson
decays like $\eta\to ee$. The current studies of $\omega$ decays 
focus on two pilot experiments and
will develop the experimental and analysis methods to process a planned high-statistics experiment
to take place before 2015. The goal is a precise determination of the $\omega\pi$ transition form
factor.

\newpage

\subsection{Recent results and perspectives on pseudoscalar mesons and form factors at BES~III}
\addtocontents{toc}{\hspace{2cm}{\sl E.~Prencipe}\par}

\vspace{5mm}

\noindent
E.~Prencipe

\vspace{5mm}

\noindent
Johannes Gutenberg University Mainz

\vspace{5mm} 
The BES~III experiment~\cite{ref0bes} is located at IHEP in Beijing (China). It is an  $e^+ e^-$ collider collecting data in a center-of-mass energy range within [2.0 $\div$ 4.6] GeV. The physics program of this project is wide and ambitious, mainly devoted to the light hadron spectroscopy. With its recent 1 billion $J/\psi$ collected and 2.9 $fb^{-1}$ integrated luminosity data collected 
at the $\psi(3770)$ energy, the BES~III project offers a unique 
opportunity to perform spectroscopy studies and measurements of 
transition form factors. The decay of $\eta$ and $\eta'$ have been 
right now under investigation in this project.

$\eta-\eta'$ mixing probes the strange quark content of light 
pseudoscalar mesons and gluon dynamics of QCD. 
Hadronic decays of $\eta'$ have garnered attention, 
in particular those to 3 pions, because they can probe isospin 
symmetry breaking. The contribution of BES~III in this sector has been:
\begin{enumerate}
\item a measurement of the branching fraction (BF) of the decay 
$\eta'\rightarrow \eta \pi^+ \pi^-$ and extrapolation of the 
Dalitz plot parameters, using a 225 million $J/\psi$ 
data sample~\cite{ref1etapp}. The hadronic decay of $\eta'$ 
is extremely valuable in studies devoted to the effect 
of the gluon component in chiral perturbation theory and the 
possible nonet of light scalars. The conclusion of this recent 
BES~III publication is that the two parameterizations used to parameterize 
the amplitude distribution (named {\em general decomposition} 
and {\em linear parameterization})~\cite{ref2pdg} are not equivalent, 
as expected. However, the parameter related to probe $C$-parity violation 
in this strong decay in both parameterizations is consistent with 0;
\item a preliminary study has been started in BES~III to calculate 
the BF($\eta'\rightarrow \pi^+ \pi^- l^+ l^-$), where $l = e, \mu$. 
Theoretical predictions make the decay 
$\eta'\rightarrow \pi^+ \pi^- e^+ e^-$ extremely more probable 
than $\eta'\rightarrow \pi^+ \pi^- \mu^+ \mu^-$. 
BES~III can now confirm the theoretical expectation and 
the results from previous experiments~\cite{ref4cleo};
\item a measurement of the transition form factors 
in the decay $e^+e^-\rightarrow e^+e^- \pi^0 / \eta / \eta'$ 
via $\gamma \gamma$ interactions has been started. 
The sample used for such a study is the data sample 
collected from BES~III at $\psi(3770)$. Feasibility studies 
show that these analyses are feasible in the range of the transfer 
momentum $Q^2$ within [0.3 $\div$ 10] GeV$^2$. The BES~III data are 
going to be in part complementary to the data already 
published from other 
experiments~\cite{ref5babar,ref5belle,ref5cleo,ref5cello} and they 
are sensitive to the best range for the hadronic light-by-light 
correction to the measurement of the muon anomaly $(g-2)_\mu$, 
which is supposed to be determined in 
[0.2 $\div$ 1.5] GeV$^2$~\cite{ref6nyffeler,ref6fisher}. 
In fact, the main motivation to perform this analysis is 
the precise measurement of $(g-2)_\mu$, as it was found to be 
by 3.6$\sigma$ deviated from the Standard Model expectation. 
The light-by-light hadronic correction to this measurement 
are essential for a precise determination of $(g-2)_\mu$ and 
test the Standard Model at low energy scale.   
\end{enumerate}

\newpage

\subsection{Measurements of the photon-meson transition form factors}
\addtocontents{toc}{\hspace{2cm}{\sl V.P.~Druzhinin}\par}

\vspace{5mm}

\noindent
V.P.~Druzhinin

\vspace{5mm}

\noindent
BINP SB RAS and Novosibirsk State University, Novosibirsk, Russia

\vspace{5mm} 
A review of the BABAR measurements~\cite{bbrff1,bbrff2,bbrff3} 
of the photon-meson 
transition form factors for the light pseudoscalar mesons $\pi^0$, $\eta$,
and $\eta^\prime$ is presented. The recent Belle measurement~\cite{belff1}
of the $\gamma^\ast\gamma\to\pi^0$ form factor is discussed and 
compared with the 
BABAR result. Belle has about two times larger systematic uncertainty 
($\sim$10\% at
$Q^2=10$ GeV$^2$). The difference between BABAR and Belle measurements is about
two systematic errors.

Further measurements of the  photon-meson transition form factors 
in existing and
future experiments are discussed. In particular, the transition form factor as
a function of $q^2$'s of both photons can be measured using BABAR data for the 
$\eta^\prime$. Plans of the two-photon single-tag measurements at VEPP-2000 at 
relatively low $Q^2$ from 0.05 t0 0.6 GeV$^2$ are presented.

\newpage

\subsection[Measurement of {$\gamma \gamma^* \rightarrow \pi^0$} transition
form factor at Belle]{Measurement of \boldmath{$\gamma \gamma^* \rightarrow \pi^0$} transition
form factor at Belle}
\addtocontents{toc}{\hspace{2cm}{\sl S.~Uehara}\par}

\vspace{5mm}

\noindent
S.~Uehara (for the Belle Collaboration)

\vspace{5mm}

\noindent
High Energy Accelerator Research Organization (KEK), Tsukuba, Japan

\vspace{5mm} 

We report a measurement of the process $\gamma \gamma^* \rightarrow \pi^0$ 
with a 759~fb$^{-1}$ data sample~\cite{pi0tff} recorded with 
the Belle detector~\cite{belle} at the KEKB 
asymmetric-energy $e^+e^-$ collider~\cite{kekb}.
The pion transition form factor, $F(Q^2)$, is measured 
for the kinematical region
$4$~GeV$^2 \simlt Q^2 \simlt 40$~GeV$^2$, where $-Q^2$ is 
the invariant mass squared of a virtual photon.

Production of a neutral pion in the single-tag two-photon
process $e^+ e^- \to (e)e \pi^0$, where $e$ and ($e$) in the final state
are a tagged and an untagged electron, respectively,
is measured, and the differential cross section in $Q^2$ is 
converted to the transition form factor~\cite{bkt,cleo}.  
A Bhabha-veto logic in our hardware trigger system~\cite{ecltrig} 
induces a significant efficiency loss in collecting the sinal
events. We have calibrated and tuned the effect in our Monte-Carlo 
simulator, using both data and Monte-Carlo samples from the radiative-Bhabha 
process with the virtual-Compton scattering configuration~\cite{rabhat}.

\begin{figure}[h]
\begin{center}
\includegraphics[width=0.6\textwidth]{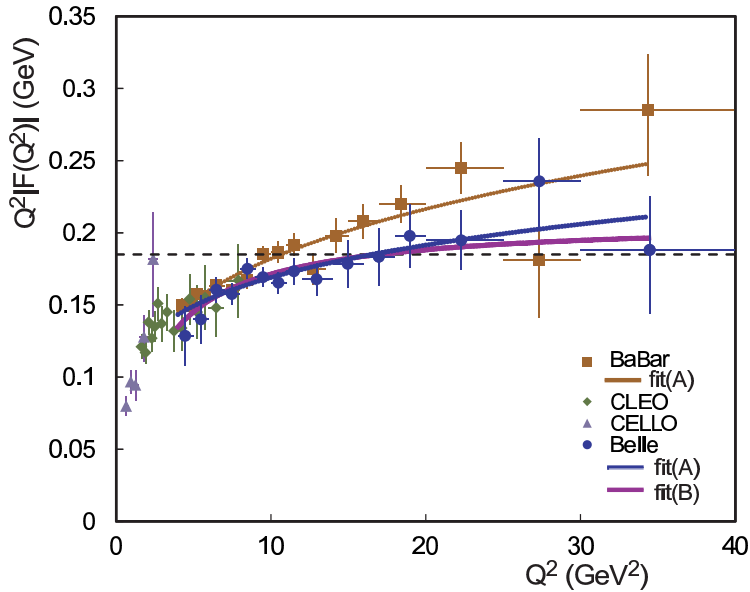}
\end{center}
\caption{Comparison of the results for the product
$Q^2|F(Q^2)|$ for the $\pi^0$ from different
experiments~\cite{pi0tff,cleo,cello,babar1}. 
The error bars are a quadratic sum of
statistical and systematic uncertainties. For the
Belle and BaBar results, only a $Q^2$-dependent
systematic-error component is included. 
The two curves denoted fit(A) 
use the BaBar parameterization, $\sim (Q^2/10$~GeV$^2)^{\beta}$,
while the curve denoted fit(B) 
uses $\sim Q^2/(Q^2+C)$~\cite{pi0tff}.
The dashed line shows
the asymptotic prediction from pQCD ($\sim 0.185$~GeV).}
\label{fig:pi0ff_aspfit}
\end{figure}

The measured values of $Q^2 |F(Q^2)|$ agree with the previous 
measurements~\cite{cello,cleo,babar1} for $Q^2 \simlt 9$~GeV$^2$.
In the higher $Q^2$ region, in contrast to BaBar, our results 
do not show a rapid growth with $Q^2$ and are closer 
to theoretical expectations~\cite{LB,bcd},
as shown in Fig.~\ref{fig:pi0ff_aspfit}.

\newpage

\subsection[Form factors and hadronic contributions to $(g-2)_\mu$ from 
Dyson-Schwinger equations]{Form factors and hadronic 
contributions to \boldmath{$(g-2)_\mu$} from 
Dyson-Schwinger equations}
\addtocontents{toc}{\hspace{2cm}{\sl C.~S.~Fischer}\par}

\vspace{5mm}

\noindent
\underline{C.~S.~Fischer}, T.~Goecke and R.~Williams

\vspace{5mm}

\noindent
JLU Giessen

\vspace{5mm} 
We review results on form factors in the Dyson-Schwinger approach and discuss
recent applications to the problem of hadronic contributions to $(g-2)_\mu$.
Based on a model for the quark-gluon interaction, the approach is capable to
reproduce experimental results on static and dynamic electromagnetic properties
of light mesons at the level of five to ten percent. Well within this error we
are also able to reproduce the result for hadronic vacuum polarization 
contributions
extracted from experiment via dispersion relations~\cite{Goecke:2011pe}. Our 
results are in good agreement with corresponding lattice calculations 
at several 
quark masses and number of active flavors. We discuss the application of our 
approach to the much debated hadronic light-by-light scattering contribution 
to $(g-2)_\mu$ and present our latest results~\cite{Fischer:2010iz,Goecke:2010if,newref}. 
We also compare our framework with the 
corresponding model approaches~\cite{newref}.

\newpage

\subsection[Hadronic contribution to $(g-2)_\mu$ from $e^+e^-$ annihilations and $\tau$ decays]{Hadronic contribution to \boldmath{$(g-2)_\mu$} from \boldmath{$e^+e^-$} annihilations and \boldmath{$\tau$} decays}
\addtocontents{toc}{\hspace{2cm}{ \sl B.~Malaescu } \par }

\vspace{5mm}

\noindent
B.~Malaescu

\vspace{5mm}

\noindent
CERN, CH--1211, Geneva 23, Switzerland

\vspace{5mm} 
A reevaluation of the hadronic contributions to the muon magnetic 
anomaly is presented. It includes the latest $\pi^{+}\pi^{-}$ cross-section 
data from the KLOE experiment, all available data from BABAR, an 
estimation of missing low-energy contributions using results on 
cross sections and process dynamics from BABAR, and an evaluation 
of the continuum contributions from perturbative QCD at four-loop order. 
The evaluation of all experimental contributions including an analysis 
of inter-experiment and inter-channel correlations is performed with 
the software package HVPTools. The new estimate features a decrease 
in the hadronic contribution with respect to earlier evaluations. 
We find that the full Standard Model prediction of the muon $g-2$ 
differs from the experimental value by $3.6\,\sigma$ ($2.4\,\sigma$) 
for the $\ee$-based ($\tau$-based) analysis. Also presented is a 
preliminary determination of $R_{e^+e^-\to{\rm hadrons}}$ 
versus the center-of-mass energy including a full evaluation of 
energy-dependent uncertainties and correlations. 
Details on the results presented can be found in 
Ref.~\cite{Davier:2010nc,Davier:2009zi,Davier:2009ag}.

\newpage

\subsection[Hadronic light-by-light scattering in the muon $g-2$:
impact of transition form factor measurements]{Hadronic light-by-light scattering in the muon \boldmath{$g-2$}: impact of transition form factor measurements}
\addtocontents{toc}{\hspace{2cm}{\sl A.~Nyffeler}\par}

\vspace{5mm}

\noindent
A.~Nyffeler 

\vspace{5mm}

\noindent
Harish-Chandra Research Institute, 
Chhatnag Road, Jhusi,
Allahabad - 211019, India

\vspace{5mm}
The calculation of the hadronic light-by-light (had.\ LbyL) scattering
contribution to the muon $g-2$ currently relies entirely on
models. Measurements of the form factors which describe the
interactions of hadrons with photons can constrain the models and
reduce the uncertainty in $a_\mu^{{\rm had.\ LbyL}}= (116 \pm 40)
\times 10^{-11}$~\cite{g-2_review}. Recently it was found in
Ref.~\cite{KLOE-2} that the KLOE-2 experiment, within one year of data
taking, could measure the decay width $\Gamma_{\pi^0\to\gamma\gamma}$
to 1\% statistical precision and the transition form factor ${\cal
F}_{\pi^0\gamma^\ast\gamma}(Q^2)$ for small space-like momenta, 
$0.01~\mbox{GeV}^2 < Q^2 < 0.1~\mbox{GeV}^2$, to 6\% statistical
precision in each bin. The impact of these measurements on estimates
of the dominant pion-exchange contribution to $a_\mu^{{\rm had.\
LbyL}}$ was also discussed in Ref.~\cite{KLOE-2}.

In the pion-exchange contribution, the form factor ${\cal
F}_{{\pi^0}^\ast\gamma^\ast\gamma^\ast}((q_1 + q_2)^2, q_1^2, q_2^2)$
with an off-shell pion with momentum $(q_1 + q_2)$ enters, see
Ref.~\cite{g-2_review} and references therein for more details. In
general, measurements of the transition form factor ${\cal
F}_{\pi^0\gamma^\ast\gamma}(Q^2) \equiv {\cal
F}_{{\pi^0}^\ast\gamma^\ast\gamma^\ast}(m_{\pi}^2, -Q^2, 0)$ are only
sensitive to a subset of the model parameters. Thus, having a good
description for ${\cal F}_{\pi^0\gamma^\ast\gamma}(Q^2)$ is only
necessary, not sufficient, to determine $a_\mu^{{\rm LbyL}; \pi^0}$.

With $\Gamma_{\pi^0\to\gamma\gamma}^{\rm PDG}$
$[\Gamma_{\pi^0\to\gamma\gamma}^{\rm PrimEx}]$ and current data for
${\cal F}_{\pi^0\gamma^\ast\gamma}(Q^2)$, the error on $a_\mu^{{\rm
LbyL}; \pi^0}$ is $\pm 4 \times 10^{-11}$ [$\pm 2 \times 10^{-11}$],
not taking into account the uncertainty related to the off-shellness
of the pion. Including the simulated KLOE-2 data reduces the error to
$\pm (0.7 - 1.1) \times 10^{-11}$~\cite{KLOE-2}. The lifetime fixes
the normalization of the transition form factor at $Q^2=0$, a source
of uncertainty in hadronic light-by-light scattering, 
which has not been considered in most evaluations.

For some models, like vector-meson dominance (VMD), which have only a
few parameters that are completely determined by measurements of the
transition form factor, this represents the total error in
$a_\mu^{{\rm LbyL}; \pi^0}$. In other models, e.g., those based on
large-$N_C$ QCD matched to the OPE, see Ref.~\cite{pion-exchange} and
references therein, there are parameters related to the off-shellness
of the pion which dominate the total error in $a_{\mu; {\rm
large-}N_C}^{{\rm LbyL}; \pi^0} = (72 \pm 12) \times 10^{-11}$.  Note
that a smaller error does not necessarily imply that a model is
better, i.e., closer to reality. Maybe the model is too
simplistic. The VMD model is known to have a wrong high-energy
behavior with too strong damping, which underestimates the
contribution, compared to the large-$N_C$ QCD model, by $15 \times
10^{-11}$, i.e., $a_{\mu; {\rm VMD}}^{{\rm LbyL}; \pi^0} \sim
57 \times 10^{-11}$.

\newpage

\subsection[Charged pion contribution to light-by-light and the muon $g-2$]{Charged pion contribution to light-by-light and the muon \boldmath{$g-2$}}
\addtocontents{toc}{\hspace{2cm}{\sl K.~Engel}\par}

\vspace{5mm}

\noindent
K.~Engel

\vspace{5mm}

\noindent
California Institute of Technology

\vspace{5mm}
In this talk, I reanalyze the charged pion contribution to the muon 
anomalous magnetic moment $a_\mu$.  Previous 
estimations~\cite{Kinoshita:1984it,Hayakawa:1995ps,Bijnens:1995cc} of this 
quantity have found a small value $(a_\mu \sim -2\pm2 \times 10^{-10})$, 
but with relatively large model dependence and a correspondingly large 
uncertainty.  Because of the sensitivity of this calculation to higher-order 
effects, it is important that the models used are as accurate as possible.  
One check on these models comes from $\chi$PT which governs the behavior 
of pions at low energies.  I present a next-to-leading (NLO) $\chi$PT 
calculation of the charged pion light-by-light amplitude and compare with a 
model based prediction~\cite{Engel:2012xb}.  I show that the NLO corrections 
can indeed be modeled using form factors for the $\gamma\pi\pi$ and 
$\gamma\gamma\pi\pi$ vertices, but that the existing models are 
incomplete, missing pion polarizability terms which show up 
in the $\gamma\gamma\pi\pi$ vertex.

In the language of resonance exchange, the current models include 
corrections due to $\rho$ exchange, but miss the pion polarizability 
corrections which are due to exchange of the $a_1$ axial vector meson.  
More complete models do exist in the literature~\cite{Ecker:1989yg} which 
contain both mesons, however, the $a_1$ exchange in these models suffers 
from poor UV behavior and cannot be used for the $a_\mu$ calculation.  
Instead, I discuss preliminary work where we model the $a_1$ exchange using 
a simple form factor approach.  Our model agrees with $\chi$PT at low 
energies, but has better UV behavior.  We have calculated numerically the 
impact of these $a_1$ exchanges on the charged pion contribution to $a_\mu$.  
I present two preliminary results which differ in their treatment of 
the $\rho$ exchanges: $a_\mu^{\rm VMD} = -6.7\times 10^{-10}$ and 
$a_\mu^{\rm HLS} = -1.6\times 10^{-10}$.  I conclude that the charged pion 
contribution appears to be enhanced by the $a_1$ exchange and may be more 
significant than previously expected.  Consequently, the uncertainty 
quoted for this quantity should also be increased.

\newpage

\subsection{Pseudoscalar Transition Form Factors @ KLOE-2 and BGO-OD} 
\addtocontents{toc}{\hspace{2cm}{\sl M.~Mascolo\\}\par}

\vspace{5mm}

\noindent
M.~Mascolo

\vspace{5mm}

\noindent
Universit\`a di Roma Tor Vergata

\vspace{5mm} 

The possibility of new pseudoscalar transition form factor measurements 
in the KLOE-2 and BGO-OD 
experiments is presented. Two-photon physics from $e^+e^-$ collisions 
is explored at DA$\Phi$NE~\cite{Amelino}; the process 
$e^+e^- \to e^+e^- \pi^0$ will be studied at KLOE-2 running the machine at the 
$\phi$ peak ($\sqrt{s} = 1.02$ GeV) thanks to new lepton tagger 
detectors~\cite{mes_masc}. In 
particular, the possibility to measure the width 
$\Gamma_{\gamma\gamma \to \pi^0}$ with a $\sim 1\%$ 
level statistical accuracy and the $\pi^0 \gamma^{*} \gamma$ form factor, 
$F(Q^2)$, at low 
invariant masses ($0.01 < Q^2 < 0.1$ GeV$^2$) of the virtual photon is 
considered (see~\cite{our_p0} for details). In the BGO-OD 
experiment~\cite{sch} pseudoscalar mesons will be photoproduced and the 
transition form factors will be measured via Dalitz decay processes 
(see~\cite{mes_mor}).

\newpage

\subsection{Meson Decay Program with Crystal Ball at MAMI}
\addtocontents{toc}{\hspace{2cm}{\sl M.~Unverzagt}\par}

\vspace{5mm}

\noindent
M.~Unverzagt

\vspace{5mm}

\noindent
Institut f\"ur Kernphysik, Johannes Gutenberg University Mainz, Germany

\vspace{5mm} The Institute for Nuclear Physics at the Johannes 
Gutenberg University in Mainz, Germany, has established a new 
Collaborative Research Centre (SFB 1044) which has been granted funding 
by the German Science Foundation (DFG) for up to 12 years. The programme of 
the SFB 1044 includes studies of electromagnetic as well as neutral $\eta$ 
and $\eta'$ decays. One of the main goals of the SFB 1044 is to significantly 
improve existing statistics from the Crystal Ball at MAMI experiment 
on the $\eta \to e^+ e^- \gamma$ decay from roughly 1500~\cite{Ber11} 
to 30,000 events and to measure the as yet unobserved 
$\eta' \to e^+ e^- \gamma$ decay. These measurements provide information 
on meson transition form factors which are also of importance in calculating 
the hadronic light-by-light contribution to the anomalous magnetic moment 
of the muon. Furthermore, the neutral decays 
$\omega \to \eta \gamma$, $\eta' \to \omega \gamma$, $\eta/\eta' \to 3 \pi^0$, and $\eta' \to \eta \pi^0 \pi^0$ will be studied to constrain 
effective field theories of QCD. With the ongoing development of a 
high-rate Time-Projection-Chamber to be installed at the center of the 
Crystal Ball the access to all charged meson decay channels will be 
improved significantly in the near future.

\newpage

\subsection{Transition Form Factors from CMD-2/CMD-3}
\addtocontents{toc}{\hspace{2cm}{\sl S.~Eidelman}\par}

\vspace{5mm}

\noindent
S.~Eidelman

\vspace{5mm}

\noindent
Budker Institute of Nuclear Physics SB RAS and Novosibirsk State University

\vspace{5mm} 
Transition form factors of the $\pi^0$ and $\eta$ mesons have been studied
at CMD-2 in a number of various processes described below.

The reactions $e^+e^- \to \pi^0\gamma$ and  $e^+e^- \to \eta\gamma$
were studied in the broad center-of-mass (c.m.) energy range from
600 to 1400~MeV using the $2\gamma$ decay mode of $\pi^0$ and 
$\eta$~\cite{akh05}. Earlier $\phi$ meson decays to $\eta\gamma$
have also been studied using the $\pi^+\pi^-\pi^0$ and  $\pi^+\pi^-\gamma$
decay modes~\cite{akh99F} and $3\pi^0$~\cite{akh01B}.

The $\phi \to \eta^{\prime}\gamma$ decay was discovered by
CMD-2 in 1997~\cite{akh97B} in the decay chain
$\phi \to  \eta^{\prime}\gamma,~\eta^{\prime} \to \eta\pi^+\pi^-,~
\eta \to 2\gamma$, the same decay modes with increased statistics
were used in~\cite{akh00B} and finally with $\eta \to \pi^+\pi^-\pi^0$
in~\cite{akh00F}. 

Studies of the conversion decays began at the $\phi$ meson
with the first observation of the $\eta e^+e^-$~\cite{akh01}
and $\pi^0 e^+e^-$~\cite{akh01C}.  Later such decays were studied
at the $\omega$ meson and searched for at the $\rho$~\cite{akh05A}.
 
The tagged $\eta$ mesons from the $\phi \to \eta \gamma$ decays have been
also used to study the Dalitz decay $\eta \to e^+e^-\gamma$, observe for the
first time the $\eta \to e^+e^-\pi^+\pi^-$ decay, and set an upper limit on the
decay $\eta \to e^+e^-e^+e^-$~\cite{akh01}.

Finally, information on transition form factors in a broad mass range
was obtained in the processes  $e^+e^- \to \pi^0\pi^0\gamma$ dominated by
the $\omega\pi^0$ intermediate mechanism~\cite{akh03B} and
$e^+e^- \to \eta\pi^+\pi^-$ dominated by the $\eta\rho$ 
state~\cite{akh00D}~(see also Ref.~\cite{ce11} for detailed references.)

In the new experiment with the CMD-3 detector it is planned to study most
of these processes with data samples two orders of magnitude higher. A
significant background caused by conversion on the detector material will
be suppressed using the new drift chamber.

\newpage

\subsection[Study of the Process $e^+e^-\to\omega\pi^0\to\pi^0\pi^0\gamma$ in c.m.\ Energy Range 1.05--2.0~GeV at SND]{Study of the Process \boldmath{$e^+e^-\to\omega\pi^0\to\pi^0\pi^0\gamma$} in c.m.\ Energy Range 1.05--2.0~GeV at SND}
\addtocontents{toc}{\hspace{2cm}{\sl L.~Kardapoltsev}\par}

\vspace{5mm}

\noindent
L.~Kardapoltsev

\vspace{5mm}

\noindent
BINP SB RAS and Novosibirsk State University, Novosibirsk, Russia

\vspace{5mm} 
A preliminary result of the  measurement of the cross section 
for the process
$e^+e^- \to \omega \pi^0 \to \pi^0\pi^0\gamma$ is presented. The
experiment has been performed with the Spherical Neutral Detector
(SND)~\cite{SND_desc,SND_desc2,SND_desc3,SND_desc4} at the
VEPP-2000 $e^+e^-$ collider~\cite{VEPP} in the energy range
1.05--2.00 GeV. The analysis is based on 27 pb$^{-1}$ collected
during 2010 and 2011. The previous measurement~\cite{SND2010} used statistics
corresponding to 5~pb$^{-1}$.

Our results are well consistent with the measurements~\cite{ppg_SND,ppg_CMD}
performed by SND and CMD-2 at the VEPP-2M collider at the energies
below 1.4 GeV,
but significantly (by 20--30\%) exceed DM2~\cite{dm2} data, the only previous
measurement at the energy above 1.4 GeV.

We also present comparison of  our data on the  cross section  with the cross
section calculated under the CVC hypothesis from the spectral function
of $\tau \to \omega \pi \nu_{\tau}$ decay measured in the CLEO
experiment~\cite{cleo-k}. A sizable difference is observed between
$e^+e^-$ and $\tau$ data.

\begin{figure}[t]
\centering
\includegraphics[width=0.8\textwidth]{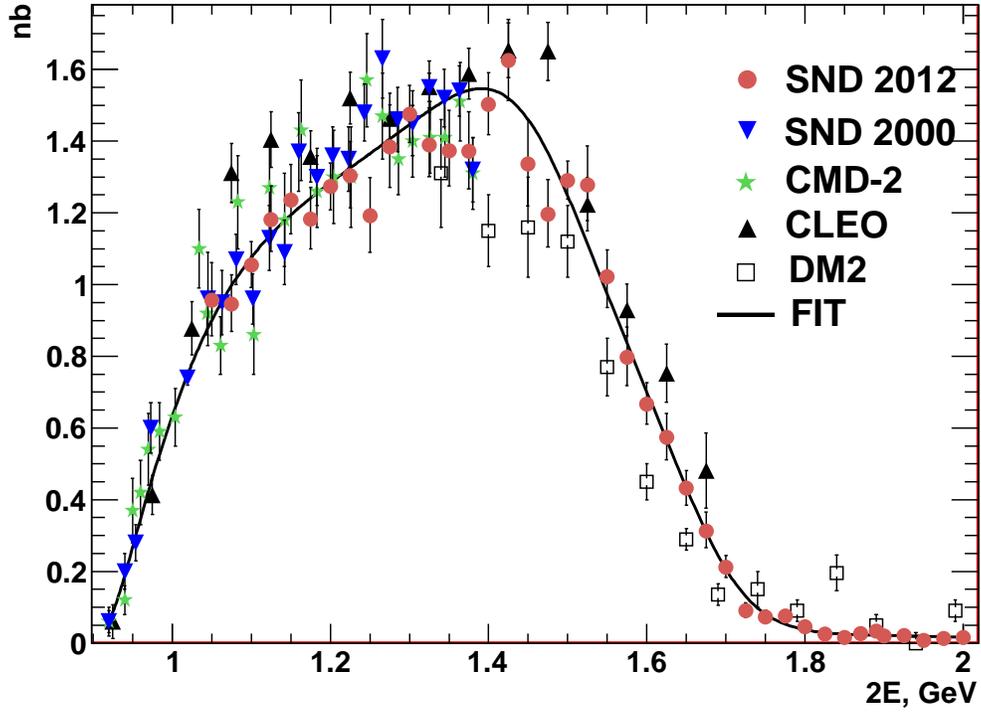}
\caption{Cross section for the 
$e^+e^- \to \omega \pi^0 \to \pi^0\pi^0\gamma$ 
process according to the SND 2012 (this work),
SND 2000~\cite{ppg_SND}, CMD-2~\cite{ppg_CMD}, CLEO~\cite{cleo-k},
and DM2~\cite{dm2} data. The curve is the result of the joint
approximation of the SND 2012 and SND 2000 data. \label{crs_ompi}}
\end{figure}

Figure~\ref{crs_ompi} shows the cross section for the process
under study according to our data, as well as SND
2000~\cite{ppg_SND}, CMD-2~\cite{ppg_CMD}, CLEO~\cite{cleo-k}, and
DM2~\cite{dm2} data. The cross section from~\cite{dm2}, measured
in the $\pi^+\pi^-\pi^0\pi^0$ channel, was recalculated using
the tabulated branching ratios of the $\omega$ meson~\cite{pdg}. Under the
assumption of vector current conservation, the CLEO cross section
was calculated using the spectral function of 
$\tau \to \omega \pi \nu_{\tau}$ decay measured in the 
CLEO experiment~\cite{cleo-k}.

\newpage

\subsection{Transition Form Factors with HADES}
\addtocontents{toc}{\hspace{2cm}{\sl W.~Przygoda}\par}

\vspace{5mm}

\noindent
W.~Przygoda

\vspace{5mm}

\noindent
Jagiellonian University, Cracow, Poland

\subsubsection*{Introduction}

\begin{sloppypar}
The High Acceptance Di-Electron Spectrometer (HADES)~\cite{spectrometer} operates at the GSI Helmholtz\-zentrum f\"ur Schwerionenforschung in Darmstadt, Germany. The scientific program covers a wide range of systematic studies of $e^+e^-$, pseudoscalar, vector mesons and baryon resonances production in nucleon-nucleon collisions ($p+p$ at 1.25, 2.2, 3.5 GeV and $d+p$ at 1.25 GeV/u), proton-nucleus reactions ($p+{\rm Nb}$ at 3.5 GeV) and heavy ion collisions (${\rm C}+{\rm C}$ at 1 and 2 GeV/u, ${\rm Ar}+{\rm KCl}$ at 1.725 GeV/u and ${\rm Au}+{\rm Au}$ at 1.25 GeV/u). Continuation of this program with pion induced reactions using secondary pion beams at GSI is currently under discussion. One of the important aspects of these studies is the investigation of $e^+e^-$ conversion processes of baryon resonances in elementary and heavy ion collisions.
\end{sloppypar}

\subsubsection*{Results from \boldmath{$p+p$} collisions at 1.25~GeV}

The beam energy was selected below $\eta$ meson production in order to favor $\Delta(1232)$ production and to investigate the $\Delta$ Dalitz decay ($\Delta \rightarrow p \gamma^* \rightarrow p e^+ e^-$). For this purpose, the $ppe^+e^-$ final state was isolated using the following conditions: $M_{ee} > 0.14~{\rm GeV}/c^2$ and missing mass of the $p e^+ e^-$ system equal to the proton mass. This decay probes the structure of the transition $\Delta(J=3/2) \rightarrow N(J=1/2) \gamma^*$ in the time-like region ($q^2 > 0$) where the description heavily depends on the model description of the respective electromagnetic form factor~\cite{waniachello,ramalhopena}. Reconstruction of the complementary channel $pp\pi^0 \rightarrow p p e^+ e^-\gamma$  allows for the identification of the $\Delta^+$ resonance as a major intermediate state and further allows to  deduce, for the first time, the branching ratio of the $\Delta$ Dalitz decay. Furthermore, recently published results on one-pion production in the $n p \pi^+$ and $p p \pi^0$ channels~\cite{beatrice} and respective comparisons with one-boson-exchange model calculations~\cite{teisdmitriev} clearly show dominance of the $\Delta^{++}$ and $\Delta^{+}$. In parallel, the partial wave analysis (inspired by~\cite{sarantsev}) has been launched with the aim of a more precise baryon resonance contribution description.

\subsubsection*{Results from \boldmath{$p+p$} at 2.2~GeV}

The higher proton beam energy used in this reaction enabled the investigation of inclusive and exclusive  $\pi^0$ and $\eta$ meson production~\cite{gumberidze,beatrice} by means of hadron and dilepton decay channels. Dalitz decays of the pion ($\pi^0 \rightarrow \gamma e^+ e^-$) and the eta meson ($\eta \rightarrow \gamma e^+ e^-$) were compared with the simulation implementing the point-like (QED) and VMD form factor. The results show negligible difference for the $\pi^0$ but clearly favors the VMD transition form-factor for the $\eta$ Dalitz decay. 

\subsubsection*{Results from \boldmath{$p+p$} at 3.5~GeV}

\begin{sloppypar}
At this beam energy higher baryon resonances with isospin $I=3/2$ and $I=1/2$ come into play and there is much higher sensitivity to investigate baryon electromagnetic time-like transition form factors (see~\cite{dohrmann,ramalhopena}). HADES results on inclusive $e^+e^-$ production~\cite{rustamov} unravel the insufficient contribution to dielectron yields of the known dilepton sources in the $0.4~{\rm GeV}/c^2< M_{ee} < 0.7~{\rm GeV}/c^2$ range. On the other hand calculations \cite{weil} using the two-component quark model (by Wan-Iachello~\cite{waniachello}) for the $\Delta(1232)$ transition form factor show better agreement with data. Yet, another possibility for better description of the data is provided by the GiBUU model~\cite{weil}, where the resonance conversion proceeds via intermediate $\rho N$ states. The exclusive analysis of $pn\pi^+$, $pp\pi^0$ and $pe^+e^-$ final states~\cite{dybczak} brings more constraints to various models. Cross sections of baryon resonances were estimated first in the hadron analysis by simultaneous fitting procedure of invariant masses ($M_{inv}^{p\pi^+}$, $M_{inv}^{n\pi^+}$, $M_{inv}^{p\pi^0}$) and angular distributions ($\cos\theta^{p\pi^+}_{CM}$, $\cos\theta^{n\pi^+}_{CM}$, $\cos\theta^{p\pi^0}_{CM}$) in several bins of the $N\pi$ invariant mass. Then the exclusive $ppe^+e^-$ was reconstructed and compared to the following models:
\begin{itemize}
\item Zetenyi-Wolf QED calculations of the resonance conversion rates~\cite{zetenyiwolf} assuming point-like photon-baryon coupling, which can serve as a lower limit of $e^+e^-$ emission. For the  $\omega/\rho$ mesons experimentally deduced cross sections are used. The calculation satisfactorily describes the $e^+e^-$ yield at the vector meson pole but cannot fully explain the measured $e^+e^-$ yield for lower $e^+e^-$ invariant mass.
\item Krivoruchenko-Martemyanov calculations~\cite{krivoruchenko} implement an extended Vector Meson Dominance (eVMD) model for the baryon conversion rates. On the contrary, we observe too large a yield at the vector meson pole and missing yield related to high mass resonances when compared to HADES data.
\item Zetenyi-Wolf calculations accomplished with Wan-Iachello form factor (only for $\Delta(1232)$)~\cite{waniachello} give a better description, but still a similar treatment of higher resonances is missing. Calculations of the transition form factors for higher $\Delta$ and $N^*$ resonances in the time-like region~\cite{ramalhopena} should help to improve the situation.
\end{itemize}
\end{sloppypar}

Yet another important result of the inclusive $e^+e^-$ analysis~\cite{rustamov} is the estimation of the upper limit of the branching ratio for the rare electromagnetic decay $\eta \rightarrow e^+e^-$, suppressed by helicity conservation. The value $< 5.6 \times 10^{-6}$ ($90\%$ CL) for the $\eta$ inclusive production in $p+p$ collisions was accepted as a new PDG entry.

\subsubsection*{Conclusions}

The HADES spectrometer provided a wealth of data from $p+p$ collisions at energies 1.25, 2.2 and 3.5 GeV, allowing for systematic studies of meson and baryon resonance production and their decays. In particular, exclusive channels with pion production allow to put constraints on the $\Delta$ and $N^*$ cross sections, which are important for a proper description of the resulting dilepton yields. As a consequence, the baryon resonance transition form factors in the time-like region can be investigated, triggering further development in the theoretical description of this interesting process.

\newpage

\subsection{Some comments on CLEO results and their interpretation}
\addtocontents{toc}{\hspace{2cm}{\sl V.~Savinov}\par}

\vspace{5mm}

\noindent
V.~Savinov

\vspace{5mm}

\noindent
University of Pittsburgh, Dept.\ of Physics and Astronomy, 3941 O'Hara St., Pittsburgh, PA 15260, USA 

\vspace{5mm} 
A personal perspective on the goals of the workshop 
and on the connections between the experiment and theory 
in the field of exclusive QCD processes is outlined. 
A brief review of the current status of hadronic light-by-light (hlbl) 
scattering contribution to the muon anomalous magnetic moment 
is presented using selected recent literature~\cite{Boughezal:2011vw,Goecke:2010if*,Prades:2009tw*}. 
It is pointed out that, in spite of impressive experimental and 
theoretical work over the past decade, the uncertainty in the 
hlbl contribution has remained the same 
(approximately $26 \times 10^{-11}$ (see, {\it e.g.}, 
Ref.~\cite{Prades:2009tw*}). 
It is remarked that, while the new experimental information on 
exclusive hadronic reactions would be welcome, there is a large theoretical 
contribution to the uncertainty that could not be trivially reduced  
using the data. The important role of strong phases in searches for 
direct CP violation and, generally, Beyond-the-Standard-Model (BSM) physics 
is emphasized, as also is the need to be able to calculate such phases 
in hadronic decays of heavy flavors with small theoretical uncertainty. 
Experimental difficulties of the CLEO analysis~\cite{Gronberg:1997fj*} 
of $\gamma^*\gamma \to \pi^0$ transition form factor are discussed. 
The trigger algorithms, data-driven measurements 
and event displays are presented. The interpretation of 
CLEO measurement in terms of the pion distribution amplitude 
is discussed and criticized. A discussion of relative merits 
of various theoretical predictions~\cite{Jakob:1994hd,Radyushkin:1996tb,Bakulev:2001pa,Bakulev:2003cs} 
is presented through the eyes of an experimentalist. 
CLEO results~\cite{Gronberg:1997fj*} 
are compared with the BaBar results~\cite{Aubert:2009mc*} and the BELLE results~\cite{Uehara:2012ag*} 
(both experiments also reported their results at this workshop). 
Recent theoretical developments~\cite{Kroll:2010bf} 
in the framework of the modified perturbative approach are highlighted. 
The importance of first-principles-based non-perturbative calculations 
is emphasized. 
Simple monopole fits to experimental data are criticized and clarified 
(such interpretations could be used primarily to simplify the comparison 
among different experiments). 
The CLEO measurement~\cite{Pedlar:2009aa} used to 
estimate~\cite{Aubert:2006cy} transition form factors 
of $\eta$ and $\eta^\prime$ 
mesons in the time-like domain is briefly discussed as also is the 
CLEO measurement~\cite{Pedlar:2005sj} of electromagnetic 
form factors of pion, kaon and proton. 
To conclude, the importance of future space-like and time-like data 
for reducing the uncertainty in the hlbl contribution to the muon 
anomalous magnetic moment 
is reiterated. It is emphasized that space-like measurements with two 
highly virtual space-like 
photons is desirable.

\newpage

\subsection{A new parameterization for the pion vector form factor}
\addtocontents{toc}{\hspace{2cm}{\sl C.~Hanhart}\par}

\vspace{5mm}

\noindent
C.~Hanhart

\vspace{5mm}

\noindent
Forschungszentrum J\"ulich

\vspace{5mm} 
A new approach to the parameterization of pion form factors is
presented and applied to the pion vector form factor for illustration. It has
the correct analytic structure, is  consistent with recent high
accuracy analyses of $\pi\pi$ scattering phase shifts at low energies, 
and, at high energies,
maps smoothly onto the well-known, successful isobar model.
With three resonances and three channels ($\pi\pi$, $4\pi$, $\omega\pi$) 
within the
model it is possible to simultaneously describe  data on $\pi\pi$ scattering,
$e^+e^-\to \pi^+\pi^-$ as well as
$e^+e^-\to$(non-$2\pi$)$_{\mbox{isovector}}$. Details of the model can be found
in Ref.~\cite{ref}.

\newpage

\subsection{High accuracy pion phase shifts and light scalar mesons}
\addtocontents{toc}{\hspace{2cm}{\sl J.~R.~Pel\'aez}\par}

\vspace{5mm}

\noindent
J.~R.~Pel\'aez

\vspace{5mm}

\noindent
Departamento de F\'{\i}sica Te\'orica II.
Facultad de CC. F\'{\i}sicas.\\ Universidad Complutense.
28040 Madrid. SPAIN

\vspace{5mm} 

We have reviewed our simple and accurate parameterizations 
of pion-pion scattering up to 1400 MeV, for the three isospin channels 
with angular momentum up to $\ell=3$. These parametrizations can be of 
interest for many hadronic processes, and form factors in particular, 
as long as they contain two pions in the final state. These fits are 
first obtained from fits to data but then are also constrained to 
satisfy three sets of dispersion relations. In particular, we impose 
three Forward Dispersion Relations up to 1400 MeV, together with three 
Roy equations and three  GKPY Equations up to 1100 GeV. The resulting 
fits, although still being simple, are nevertheless able to satisfy 
the analyticity constraints while describing the data. The method is 
sufficiently powerful to  disentangle a longstanding controversy 
between two sets of data on the inelasticity of the scalar-isoscalar 
wave around the $f_0(980)$ resonance. 

In addition, the use of the partial wave dispersion relations in the 
form of Roy Equations, and particularly in the form of GKPY equations, 
also allow us to provide a precise determination of the $f_0(500)$ and 
$f_0(980)$ poles and residues.

This talk was based on our recent works 
in~\cite{GarciaMartin:2011cn,GarciaMartin:2011jx},
to which we refer the reader for further details and references.

\newpage

\subsection[Roy--Steiner equations for $\gamma\gamma\to\pi\pi$]{Roy--Steiner equations for \boldmath{$\gamma\gamma\to\pi\pi$}}
\addtocontents{toc}{\hspace{2cm}{\sl M.~Hoferichter}\par}

\vspace{5mm}

\noindent
\underline{M.~Hoferichter}$^{1,2}$, D.~R.~Phillips$^{2}$ and C.~Schat$^{2,3}$

\vspace{5mm}

\noindent
$^{1}$ Helmholtz-Institut f\"ur Strahlen- und Kernphysik (Theorie) and Bethe Center for Theoretical Physics, Universit\"at Bonn\\
$^{2}$ Institute of Nuclear and Particle Physics and Department of Physics and Astronomy, Ohio University\\
$^{3}$ CONICET - Departamento de F\'{\i}sica, FCEyN, Universidad de Buenos Aires

\vspace{5mm} 

We present a system of Roy--Steiner equations for $\gamma\gamma\to\pi\pi$ 
that respects analyticity, unitarity, gauge invariance, and crossing 
symmetry~\cite{HPS11}. In general, the derivation of these equations 
proceeds similarly to the construction of Roy equations for $\pi\pi$ 
scattering~\cite{Roy}, however, starting from hyperbolic dispersion 
relations~\cite{HS_73} to account for the fact that crossing symmetry 
intertwines $\gamma\pi\to\gamma\pi$ and $\gamma\gamma\to\pi\pi$. 
Assuming elastic unitarity, the equations for the $\gamma\gamma\to\pi\pi$ 
partial waves can be solved by a single-channel Muskhelishvili--Omn\`es 
representation with finite matching point~\cite{MuskhelishviliOmnes,piK}. 
To suppress the dependence of observables on high-energy input, we also 
consider once- and twice-subtracted versions of the equations, and 
identify the subtraction constants with dipole and quadrupole pion 
polarizabilities. We present the results for low-energy cross sections 
using $\pi\pi$ phase shifts from~\cite{madrid,bern} and pion 
polarizabilities from~\cite{GIS,GM_10}. In the same way as $\pi\pi$ 
Roy equations may be used to determine rigorously the pole parameters 
of the $\sigma$ resonance~\cite{CCL}, Roy--Steiner equations for 
$\gamma\gamma\to\pi\pi$ give access to its coupling to two photons. 
With input for the polarizabilities from~\cite{GIS}, we find a partial 
width $\Gamma_{\sigma\gamma\gamma}=(1.7\pm 0.4)\,{\rm keV}$. 
Additional information on the pion polarizabilities would further 
constrain the $\sigma\gamma\gamma$ coupling.

The careful analytic continuation required to obtain the residue 
at the $\sigma$ pole shows that calculating a simple 
``$\sigma$-pole'' contribution to hadronic light-by-light scattering 
amounts to an uncontrolled approximation of the amplitudes relevant 
for the charged-pion loops. In contrast, the framework presented in 
this talk, although constructed for on-shell photons in the first place, 
might be valuable to extend low-energy models for the charged-pion-loop 
contribution to higher energies~\cite{Engel}.

\newpage

\subsection[A Dispersive Treatment of $K_{\ell4}$ Decays]{A Dispersive Treatment of \boldmath{$K_{\ell4}$} Decays}
\addtocontents{toc}{\hspace{2cm}{\sl P.~Stoffer}\par}

\vspace{5mm}

\noindent
G.~Colangelo$^{(a)}$,
E.~Passemar$^{(b)}$,
\underline{P.~Stoffer}$^{(a)}$

\vspace{5mm}

\noindent
$^{(a)}$ Albert Einstein Center for Fundamental Physics, Institute for Theoretical Physics, University of Bern, Sidlerstrasse 5, CH-3012 Bern, Switzerland \\
$^{(b)}$ Theoretical Division, Los Alamos National Laboratory, Los Alamos, NM 87545, USA

\vspace{5mm} 

$K_{\ell4}$ is for several reasons an especially interesting decay 
channel of $K$ mesons:
it allows an accurate measurement of a combination of
$S$-wave $\pi\pi$ scattering lengths, one form factor of the decay is
related to the chiral anomaly and the decay is the best source for the
determination of various low energy constants of ChPT.

We present a dispersive approach to $K_{\ell4}$ decays, which  fully
takes  into account final state rescattering effects.
Fits to the data of the E865~\cite{Pislak2003,Pislak2010} and 
NA48/2~\cite{Batley2008} experiments and results of the matching 
to ChPT are shown.

Details on the dispersion relation will be available in a 
forthcoming publication~\cite{Colangelo2012}.

\newpage

\subsection[$\eta,\eta' \to \pi^+ \pi^- \gamma$ -- A model-independent approach]{\boldmath{$\eta,\eta' \to \pi^+ \pi^- \gamma$} -- A model-independent approach}
\addtocontents{toc}{\hspace{2cm}{\sl A.~Wirzba}\par}

\vspace{5mm}

\noindent
A.~Wirzba

\vspace{5mm}

\noindent
Forschungszentrum J\"ulich

\vspace{5mm} 
We report on a new,  model-independent method to analyze radiative 
decays of mesons to a vector, isovector pair of pions of invariant 
mass squared below the first significant
$\pi\pi$ threshold in the vector channel~\cite{Stollenwerk}.
It is based on a combination of chiral perturbation
theory and dispersion theory. This allows for a controlled inclusion 
of resonance
physics without the necessity to involve vector meson dominance explicitly.
 In particular, the method is applied to an analysis of the 
reactions $\eta\to
 \pi^+\pi^-\gamma$ and  $\eta^\prime\to
 \pi^+\pi^-\gamma$.  The pertinent decay amplitude factorizes into a 
universal non-perturbative
 part, the pion isovector form factor, and a reaction-specific 
perturbative part, which is governed by a fit to the
 branching ratio and the spectrum.
 The method allows experimentalists to parameterize and compare 
decay data via two parameters only,
 whereas it allows theorists to relate different radiative decays, 
{\it e.g.}, transition form factors, branching
 ratios, slope parameters etc.

\newpage

\subsection[The $\omega/\phi\to\pi^0\gamma^*$ transition form factors in dispersion theory]{The \boldmath{$\omega/\phi\to\pi^0\gamma^*$} transition form factors in dispersion theory}
\addtocontents{toc}{\hspace{2cm}{\sl S.~P.~Schneider}\par}

\vspace{5mm}

\noindent
S.~P.~Schneider

\vspace{5mm}

\noindent
Helmholtz-Institut f\"ur Strahlen- und Kernphysik (Theorie) and Bethe Center for Theoretical Physics, Universit\"at Bonn, Germany

\vspace{5mm}
We present a study of the $\omega\to\pi^0\gamma^*$ and 
$\phi\to\pi^0\gamma^*$ electromagnetic transition form factors 
based on dispersion theory, 
a framework that is derived from fundamental principles like 
unitarity, analyticity, and crossing symmetry.
The analysis relies solely on the input from a previous dispersive 
calculation of the corresponding three-pion decays~\cite{SPS:ref1} 
and the pion vector form factor.
While our calculation of the transition form factor exhibits a clear 
improvement over naive vector meson dominance, we are---similarly to 
other theoretical approaches~\cite{SPS:ref2}---not able to account 
for the steep rise towards the end of the physical region found in 
recent measurements of the $\omega\to\pi^0\mu^+\mu^-$ decay 
spectrum by the NA60 collaboration~\cite{SPS:ref3,SPS:ref4}.
If such a deviation from the dispersion-theoretical approach has 
physical significance, it should be found in the related transition 
form factor of the next-lightest isoscalar vector meson $\phi$,
where the accessible dilepton invariant mass is somewhat larger and 
encompasses the region of the $\rho$ resonance. We thus
strongly encourage an experimental investigation of the 
Okubo--Zweig--Iizuka-forbidden $\phi\to\pi^0\ell^+\ell^-$ decays 
in order to understand these strong deviations, while additional
information on the $\omega\to\pi^0\gamma^*$ transition form factor 
to back up the NA60 data would be most welcome. 
Details of our approach can be found in Ref.~\cite{SPS:ref5}.

\newpage

\subsection{A Rational Approach to Meson Transition Form Factors}
\addtocontents{toc}{\hspace{2cm}{\sl P.~Masjuan}\par}

\vspace{5mm}

\noindent
P.~Masjuan

\vspace{5mm}

\noindent
Departamento de F\'isica Te\'orica y del Cosmos and CAFPE, Universidad de Granada, E-18071 Granada, Spain

\vspace{5mm} 

The measured $\gamma^*\gamma \rightarrow \pi^0$ transition form factor 
in the space-like region by the CLEO, CELLO, BABAR, and Belle 
collaborations are analyzed using the method developed in 
Ref.~\cite{PM:ref1} which is based on the mathematical theory of 
Pad\'{e} Approximants. The method provides a good and systematic 
description of the low energy region exemplified here with the extraction 
of the slope $a_{\pi}$ and curvature $b_{\pi}$ of the form factor in 
a model-independent way. Their impact on the pion exchange contribution 
to the hadronic light-by-light scattering part of the anomalous magnetic 
moment $a_{\mu}$ is also discussed. The main results and the details of 
the method for the transition form factor can be found in 
Ref.~\cite{PM:ref2}.

\newpage

\subsection{Radiative transitions of vector mesons}
\addtocontents{toc}{\hspace{2cm}{\sl S.~Ivashyn}\par}

\vspace{5mm}

\noindent
S.~Ivashyn

\vspace{5mm} 
\noindent
NSC ``Kharkov Institute of Physics and Technology'', Kharkov, Ukraine

\vspace{5mm} 
The conversion decay of a light vector resonance $V$
into a light pseudoscalar meson $P$ and a lepton pair $l^+l^-$
can be used as an electromagnetic probe of the non-perturbative 
phenomena in the odd-intrinsic-parity sector at low energies.
The classic review on the radiative decays of mesons~\cite{Landsberg:1986fd-SI} 
dates back to 1986 and discusses the experimental aspects of 
$V \to P l^+l^-$ measurements. The basic formulae which are typically used
in the extraction of the transition form factor from data are also presented therein.
The cross-channel processes $l^+l^- \to P V$ and $P \to V l^+l^-$
allow for the extraction of the transition form factor 
in yet another kinematical regions and  
are also considered in~\cite{Landsberg:1986fd-SI}.
Since that time, new measurements of the transition form factors
have been performed and new theoretical descriptions of the form factors
developed.
Also the vector meson radiative transitions have been studied in
Lattice QCD~\cite{Woloshyn:1986pk-SI,Crisafulli:1991pn-SI}.

The $V P \gamma^\ast$ transition amplitude reads
\begin{equation*}
      {\mathcal{M}}(V^{(\alpha)} \; P(k) \; \gamma^{*(\beta)}(q)) 
      = e \; g_{V P \gamma} \; F_{VP\gamma^*}(q^2) 
      \; \varepsilon^{\mu\nu\alpha\beta}
      q_\mu k_\nu \ 
\end{equation*}   
in terms of the radiative transition form factor $F_{VP\gamma^*}(q^2)$
with $q^2=M_{l^+l^-}^2$ being the virtuality of the photon coupled to 
the $VP$ vertex;
$\varepsilon^{\mu\nu\alpha\beta}$ is the Levi-Civita symbol,
$q$ and $k$ are 4-momenta of the photon and the pseudoscalar meson.
The normalization of the form factor is usually chosen such that
$F_{VP\gamma^*}(0) \equiv 1$ and the interaction strengths are given by 
the constants $g_{V P \gamma}$.
The radiative decays of vector mesons into a pseudoscalar meson and a real photon
provide an access to the values of the $g_{V P \gamma}$ constants
via the decay widths:
\begin{equation*}
      \Gamma(V \to P\gamma) = 
      \left| g_{V P \gamma} \right|^2
      \frac{e^2}{96\pi}\left(\frac{M_V^2 - m_P^2}{M_V}\right)^3   \ .   
\end{equation*}
The following constants are available from vector meson decays:
$g_{\rho\pi\gamma}$,  
$g_{\omega\pi\gamma}$,  
$g_{\phi\pi\gamma}$,
$g_{\rho\eta\gamma}$, 
$g_{\omega\eta\gamma}$, 
$g_{\phi\eta\gamma}$,
$g_{\phi\eta^\prime\gamma}$.
From the pseudoscalar meson decay widths,
\begin{equation*}
      \Gamma(P \to V\gamma) =
      \left| g_{V P \gamma} \right|^2
      \frac{e^2}{32\pi}\left(\frac{M_V^2 - m_P^2}{M_V}\right)^3      \ ,
\end{equation*}
one can access the following constants:
$g_{\rho\eta^\prime\gamma}$,
$g_{\omega\eta^\prime\gamma}$.
Up to date, all the above mentioned decays have been 
studied experimentally~\cite{Nakamura:2010zzi-SI}.
The pattern of $g_{V P \gamma}$ values
gives us the primary information on the flavor $SU(3)$ symmetry realization
in the light meson sector and, in particular, on the meson mixing.
Hence a big attention from the theory side to the radiative 
decays~\cite{Pham:2010sr-SI,Escribano:2005qq-SI,Azimov:2002nh-SI,Wang:2001kd-SI,Spalinski:1989fh-SI}.

Among the existing theoretical descriptions of the $VP\gamma$ transitions 
there are
the ``traditional'' vector meson dominance (VMD) ansatz~\cite{O'Donnell:1981sj-SI};
the effective Lagrangian approaches to a (modified) VMD~\cite{Kaiser:1990yf-SI,Bramon:1994pq-SI,Klingl:1996by-SI};
the extended Nambu-Jona-Lasinio (eNJL) model~\cite{Prades:1993ys-SI};
the hidden local symmetry Lagrangian approach 
(HLS)~\cite{Hashimoto:1996hsa-SI,Hashimoto:1996ny-SI,Benayoun:2007cu-SI,Benayoun:2009im-SI};
the linear sigma model with constituent quarks~\cite{Napsuciale:2002nk-SI};
quark models~\cite{Becchi:1965zz-SI,Bramon:2000fr-SI}
and others.
The state of the art in the field of $VP\gamma$ transitions
with the real photon is that in many theoretical approaches one can have
a simultaneous fit of the widths $\Gamma(V \to P\gamma)$ 
with a reasonable quality.

There were also numerous theoretical studies of 
the $V P \gamma^\ast$ transitions with the virtual photon,
e.g., the HLS calculations~\cite{Hashimoto:1996ny-SI},
the eNJL model approach~\cite{Prades:1993ys-SI},
an extended VMD approach~\cite{Faessler:1999de-SI},
the Dyson-Schwinger equation studies~\cite{Maris:2002mz-SI},
the light-cone quark model prediction~\cite{Qian:2008px-SI}.
The most recent theoretical advances in the modeling of the 
$V P \gamma^\ast$ transition form 
factors~\cite{Schneider:2012ez-SI,Ivashyn:2011hb-SI,Terschlusen:2011pm-SI,Terschluesen:2010ik-SI}
were partly motivated by a drastic discrepancy between
a novel CERN SPS NA60 experiment data~\cite{:2009wb-SI,Usai:2011zz-SI}
and a naive VMD ansatz prediction for
the $\omega\to\pi \gamma^\ast$ transition form factor.
The published experimental information on the 
$V \to P \gamma^\ast$ is the following:
 Lepton-G (Serpukhov): {$\omega\to\pi\mu^+\mu^-$}~\cite{Dzhelyadin:1980tj-SI};
 CMD-2 (Novosibirsk):  {$\omega\to\pi e^+e^-$}~\cite{Akhmetshin:2005vy-SI};
 SND (Novosibirsk):  {$\phi\to\eta e^+e^-$}~\cite{Achasov:2000ne-SI},
        {$\omega\to\pi e^+e^-$}~\cite{Achasov:2008zz-SI};
 NA60 (CERN): {$\omega\to\pi\mu^+\mu^-$}~\cite{:2009wb-SI,Usai:2011zz-SI}.
The form factors are extracted from the
decay line shape using the formula~\cite{Landsberg:1986fd-SI}
\begin{align*}
\frac{d\,\Gamma(V \to P \mu^+\mu^-)}{d\,Q^2} &=
\frac{\alpha}{3\pi} 
\frac{\Gamma(V \to P \mu^+\mu^-)}{Q^2}
\left( 1 + \frac{2 m_\mu^2}{Q^2} \right)
\sqrt{ 1 - \frac{4 m_\mu^2}{Q^2} }  \\
& \times 
\left( \left( 1 + \frac{Q^2}{M_V^2 - m_P^2}\right)^2  
       - \frac{4 M_V^2 Q^2}{(M_V^2 - m_P)^2}
       \right)^{3/2}
\left| F(Q^2) \right|^2 \ .
\end{align*}

A promising complementary process is $e^+e^-\to PV$
which allows to study the $V P \gamma^\ast$ form factors
in the region of time-like photon virtuality above the $PV$ threshold.
For example, the modeling of the process $e^+e^-\to \omega\pi^0$
has been considered in the NJL model~\cite{Arbuzov:2010xi-SI},
non-relativistic quark model~\cite{Kittimanapun:2008wg-SI},
an effective Lagrangian approach~\cite{Li:2008xm-SI}, etc.
The main experimental information on $e^+e^-\to \omega\pi^0$
is the following:
SND data near the phi meson~\cite{Achasov:1999wr-SI},
from the $\omega\pi^0$ threshold up to $1.4$~GeV~\cite{Achasov:2000wy-SI}
and
in the energy range $1.1-1.9$~GeV~\cite{Achasov:2012zza-SI};
there is also
CMD-2 data in the energy range $0.92-1.38$~GeV~\cite{Akhmetshin:2003ag-SI}.
The form factors can be extracted from the
cross section using the formula~\cite{Landsberg:1986fd-SI}
\begin{equation*}
      \sigma_{e^+e^- \to VP}(s) =
      \left| g_{V P \gamma} F_{VP\gamma^*}(s) \right|^2
      \frac{e^4}{12\pi s^2}
      4\sqrt{s} 
      \left(\frac{(s - (M_V + m_P)^2)(s - (M_V - m_P)^2)}{4s}\right)^{3/2} \ .
\end{equation*}

\newpage

\subsection{Decays with Vector Mesons}
\addtocontents{toc}{\hspace{2cm}{\sl C.~Terschl\"usen}\par}

\vspace{5mm}

\noindent
\underline{C.~Terschl\"usen}, B.~Strandberg, S.~Leupold

\vspace{5mm}

\noindent
Uppsala University

\vspace{5mm} 

While the vector-meson dominance (VMD) model describes some reactions 
as, e.g., the pion and eta form factors (pion and eta, respectively, 
coupled to two real or virtual photons), it fails to describe others 
as, e.g., the omega-pion form factor (omega coupled to pion and 
virtual photon)~\cite{NA60}. For calculating decays including vector 
mesons, we use a new counting scheme treating both the light pseudoscalar 
and the light vector-meson nonet on the same footing~\cite{cs1,cs2}. 
Using this counting scheme, one can describe the omega-pion form factor 
very well~\cite{omegapaper,cs2}. Additionally, the decay width for the 
decay of omega into three pions agrees very well with the experimental 
data~\cite{omega3pi}. If one in addition includes the Wess--Zumino--Witten 
action as the leading-order contribution from chiral perturbation theory, 
the scattering $e^+$ $e^-$ into three pions is well 
described~\cite{scattering}. Furthermore, the results for the pion- and 
eta-transition form factors are in numerical agreement with 
VMD~\cite{pi-eta-form-factors,bormio-proceedings} for one real and one 
virtual photon. If both photons are virtual, we find significant 
deviations from VMD.

\newpage

\subsection[Meson form factors in amplitudes for three-body $B$ decays]{Meson form factors in amplitudes for three-body \boldmath{$B$} decays}
\addtocontents{toc}{\hspace{2cm}{\sl L.~Le\'sniak}\par}

\vspace{5mm}

\noindent
L.~Le\'sniak

\vspace{5mm}

\noindent
Henryk Niewodnicza\'nski Institute of Nuclear Physics,
Krak\'ow

\vspace{5mm}
A summary of two phenomenological analyses of the rare three-body $B$ 
decays is given.
We study CP violation and the contributions of the  
strong interactions between the pion and kaon pairs in the
$B^{\pm} \rightarrow \pi^+ \pi^- \pi^{\pm}$ and 
$B^{\pm} \rightarrow K^+ K^- K^{\pm}$ decays
\cite{LL:ref1,LL:ref2}. The $B$-decay amplitudes are derived in the QCD 
factorization approach supplemented with the inclusion of the long 
distance $\pi^+ \pi^-$
and $K^+ K^-$ interactions. The form factors, corresponding to the
transitions from $B$ to light mesons or resonances and to 
transitions between light
mesons, are important components of the decay amplitudes. 
A unitary model is constructed 
for the scalar non-strange and strange form factors in which 
three scalar resonances 
$f_0(500),f_0(980)$ and $f_0(1400)$ are naturally incorporated. 
These form factors satisfy the
constraints coming from the chiral perturbation theory. 

Using our model the Dalitz plot analyses
 of the  $B^{\pm} \rightarrow \pi^+ \pi^- \pi^{\pm}$ and 
$B^{\pm} \rightarrow K^+ K^- K^{\pm}$ decays can be improved by reducing 
a number of fitted free parameters.
   The model can be extended to study CP violation in other charmless 
decays and to analyze new 
high-statistics data from Belle, BaBar, LHCb and from future super-B 
factories. Application of 
a similar approach to the charged and neutral $B$ decays 
into $K\pi^+ \pi^-$ is described in 
\cite{LL:ref3}.

\newpage

\subsection{ChPT calculations of form factors}
\addtocontents{toc}{\hspace{2cm}{\sl K.~Kampf}\par}

\vspace{5mm}

\noindent
K.~Kampf

\vspace{5mm}

\noindent
Charles University, Prague

\vspace{5mm} 
An overview on Chiral perturbation theory calculations of form factors was
presented~\cite{Bijnens:2006zp}. The main focus was given on the form factors
related to the lightest meson, pion,
namely: pion decay constant~\cite{pdc}, pion vector and scalar form factor,
radiative pion decay and transition form factor.
Finally, due to its importance for the basis of Chiral perturbation theory, also
an overview of the kaon decay form factors, namely those related to $K_{l4}$
decay, was given.
A pure calculation within the effective theory can be extended using further
methods: resonance chiral theory~\cite{res}, leading logarithm 
calculations~\cite{ll}, 
resummed chiral perturbation theory, etc.

\newpage

\subsection{Pion-photon transition form factor at the crossroads}
\addtocontents{toc}{\hspace{2cm}{\sl N. G. Stefanis}\par}

\vspace{5mm}

\noindent
N. G. Stefanis

\vspace{5mm}

\noindent
Institut f\"{u}r Theoretische Physik II,
Ruhr-Universit\"{a}t Bochum,
D-44780 Bochum, Germany

\vspace{5mm}
Using light-cone sum rules, we retrofit the CELLO, CLEO, BaBar, and 
the brand-new Belle data for the processes $\gamma^*\gamma\to \pi^0$
and $\gamma^*\gamma\to \eta$, and $\gamma^*\gamma\to \eta'$ 
beyond the level of the first two Gegenbauer coefficients in the pion
distribution amplitude $\varphi_{\pi}(x,Q^2)$~\cite{BMPS11}.
The next-to-leading order QCD perturbative contribution and the twist-four
term are taken into account explicitly, while the next-to-next-to-leading
order radiative correction~\cite{MS09} and the twist-six term are included 
by means of theoretical uncertainties.
Evolution of the pion distribution amplitude is also taken into account 
at the next-to-leading-order level.
We claim that the observed bifurcation of the Belle and the BaBar data 
above 10~GeV$^2$ is artificial, arguing that the BaBar data which show
an auxetic behavior with $Q^2$ are incompatible with the standard 
framework of QCD, while the Belle data saturate and scale with $Q^2$ as
predicted by QCD.  
We also examine the predictions for $(3/5)Q^{2}F^{\gamma^*\gamma n}(Q^2)$ 
for the state 
$|n\rangle =(1/\sqrt{2})(|u\bar{u}\rangle + |d\bar{d}\rangle)$,
extracted from the CLEO and the BaBar data on the $\eta$ and $\eta'$ 
using the quark-flavor mixing scheme.
We find very good agreement between our predictions and both data sets 
which implies that the shapes of the $\pi^0$ and the non-strange component 
$|n\rangle$ of the $\eta$ and $\eta'$ mesons are quite similar to each
other with little room for a significant flavor-symmetry violation in the 
pseudoscalar meson sector of QCD~\cite{Ste_LC2011,SBMP11}.
These distribution amplitudes are best described by the two-parameter
model of~\cite{BMS01}, which was extracted from QCD sum rules with 
nonlocal condensates, and is double-humped but endpoint suppressed. 
The Chernyak-Zhitnitsky and the asymptotic distribution amplitudes were
found before~\cite{BMS02,BMS03,BMS05lat} to be incompatible with the CLEO
data at the level of $4\sigma$ and $3\sigma$, respectively. 
Because the theoretical approaches to describe the antithetic behavior
of the Belle and the BaBar data above $10$~GeV$^2$ are hardly comparable 
to each other, we consider an interpretation of these data against some 
common standard rather questionable~\cite{BMPS12}.

\newpage

\subsection{Light pseudoscalar meson decays into lepton pair}
\addtocontents{toc}{\hspace{2cm}{\sl A.~E.~Dorokhov}\par}

\vspace{5mm}

\noindent
A.~E.~Dorokhov

\vspace{5mm}

\noindent
Bogoliubov Laboratory of Theoretical Physics, JINR, 141980 Dubna, Russia, and \\
Bogoliubov Institute of Theoretical Problems of Microworld, Lomonosov Moscow State University, Moscow 119991, Russia

\vspace{5mm}
A recent publication by the KTeV Collaboration of the branching ratio 
for the $\pi^0 \to e^+e^-$ decay~\cite{Abouzaid:2006kk} renews the 
interest to the old problem of the light pseudoscalar meson decays into 
a lepton pair. From theoretical point of view the real part of the decay 
amplitude consists of the structure-independent (dominant) and the 
structure-dependent (small) parts. The first part contains the large 
logarithms of the ratio $m_e/m_\pi$.
The second one starts from the unknown Low Energy Constant $\chi_P$. 
This constant, besides the above process, enters amplitudes of another 
observed process: the decay $\eta \to \mu^+\mu^-$~\cite{Abegg:1994wx}, 
the hadronic part of the light-by-light scattering to muon 
$g-2$~\cite{Bennett:2006fi}. In~\cite{Dorokhov:2007bd} it was shown that 
the constant $\chi_P$  can be defined as the inverse moment of the pion 
transition form factor and rather precisely extracted by using 
CLEO data~\cite{Gronberg:1997fj}. 
Then, the branching for the $\pi^0 \to e^+e^-$ decay can be predicted 
in a model-independent way. It turns out that the theoretical 
prediction deviates from the experimental number by 3.3 $\sigma$.

At present, the principal point is to perform new measurements 
of the light pseudoscalar meson decays into a lepton pair as a test of 
the Standard Model. Besides the pion decay, the most interesting modes 
are the muon modes of the $\eta$ and $\eta'$ meson decays, where 
theoretically not only lower, but also upper bounds are 
predicted~\cite{Dorokhov:2007bd,Dorokhov:2009xs}. Recently, new 
experimental upper bounds on the decay $\eta \to e^+e^-$ were 
obtained~\cite{Berlowski:2008zz*,HADES}. Present and future 
measurements~\cite{HADES,BESIII,KLOE,CMD}  of the pseudoscalar meson 
transition form factors at low momenta (WASA-at-COSY, KLOE-2, BESIII, 
CMD, HADES) will help to define the constant $\chi_P$ more precisely.

The confirmation of the deviation between experiment and theory 
will indicate  existence of hypothetical Dark Matter particles with 
low mass (of order 10--100 MeV) (see~\cite{Kahn:2007ru}).

\newpage

\subsection{The hadronic light-by-light contribution to the muon anomalous magnetic moment}
\addtocontents{toc}{\hspace{2cm}{\sl J.~Bijnens}\par}

\vspace{5mm}

\noindent
J.~Bijnens

\vspace{5mm}

\noindent
Department of Astronomy and Theoretical Physics, Lund University\\
S\"olvegatan 14A, SE 22362 Lund, Sweden

\vspace{5mm} 
The hadronic light-by-light contribution to the muon
anomalous magnetic moment will become a limiting factor in the 
accuracy of the theory prediction in the Standard Model in the near future.
A fairly recent update of the situation can be found in the presentations
at the INT workshop in February-March 2011~\cite{INT}.

There are at present three main calculations of this quantity,
BPP~\cite{BPP}, HKS~\cite{HKS} and MV~\cite{MV}, all following the separation
scheme of~\cite{deRafael}. We have recently studied the contributions
of the different momentum regions in more detail. This work was started in
\cite{BP} and for the pion-loop we now understand~\cite{BZ} the large 
difference in the results in the full VMD model~\cite{BPP} and the
hidden local symmetry model~\cite{HKS}. We have also studied the influence of
$L_9$ and $L_9+L_{10}$ which was recently suggested to be important
\cite{EPR}. This method also allows to clearly see where the quark-loop
contributes~\cite{BZ}.

\newpage

\section{List of participants}

\begin{flushleft}
\begin{itemize}
\item Benayoun, Maurice, LPNHE des Universit\'es Paris VI et Paris VII, {\tt benayoun@in2p3.fr}
\item Ber{\l}owski, Marcin, National Centre for Nuclear Research, Warsaw, Poland {\tt berlowsk@fuw.edu.pl}
\item Bhatt, Himani, IIT Bombay and Forschungszentrum J\"ulich, {\tt himani2711@gmail.com}
\item Czerwi\'nski, Eryk, Jagiellonian University, {\tt eryk.czerwinski@uj.edu.pl}
\item Czy\.z, Henryk, Inst.\ of Physics, Univ.\ of Silesia, {\tt henryk.czyz@us.edu.pl}
\item Dorokhov, Alexander E., Bogoliubov Laboratory of Theoretical Physics, JINR, 141980 Dubna, Russia, {\tt dorokhov@theor.jinr.ru}
\item Druzhinin, Vladimir P., 	Budker Institute of Nuclear Physics SB RAS and
Novosibirsk State University, Novosibirsk,  {\tt v.p.druzhinin@inp.nsk.su}
\item Eidelman,	Simon,	Budker Institute of Nuclear Physics SB RAS and
Novosibirsk State University, Novosibirsk, {\tt eidelman@mail.cern.ch}
\item Engel Kevin, California Institute of Technology, {\tt kte@caltech.edu}
\item Escribano, Rafel, Universitat Autonoma de Barcelona, {\tt Rafel.Escribano@ifae.es}
\item Fischer, Christian~S., JLU Giessen, {\tt christian.fischer@theo.physik.uni-giessen.de}
\item Goldenbaum, Frank, Forschungszentrum J\"ulich, {\tt f.goldenbaum@fz-juelich.de}
\item Goswami, Ankita, IIT Indore, {\tt ankitaorg4@gmail.com}
\item Gumberidze, Malgorzata, IPNO, {\tt sudol@ipno.in2p3.fr}
\item Hanhart, Christoph, Forschungszentrum J\"ulich, {\tt c.hanhart@fz-juelich.de}
\item Hodana, Malgorzata, Jagiellonian University, {\tt m.hodana@gmail.com}
\item Hoferichter, Martin, Helmholtz-Institut f\"ur Strahlen- und Kernphysik (Theorie) and Bethe Center for Theoretical Physics, Universit\"at Bonn, {\tt hoferichter@hiskp.uni-bonn.de}
\item Ivashyn, Sergiy, ITP NSC KIPT, Kharkov, {\tt s.ivashyn@gmail.com}
\item Jansen, Karl, NIC, DESY, Platanenallee 6 , 15738 Zeuthen, {\tt Karl.Jansen@desy.de}
\item Jegerlehner, Fred, DESY Zeuthen/Humboldt University Berlin, {\tt fjeger@physik.hu-berlin.de}
\item Kampf, Karol, Charles University, Prague, {\tt kampf@ipnp.troja.mff.cuni.cz}
\item Kardapoltsev, Leonid, 	Budker Institute of Nuclear Physics SB RAS and
Novosibirsk State University, Novosibirsk, {\tt l.kardapoltsev@gmail.com}
\item Khan, Farha Anjum, Forschungszentrum J\"ulich, {\tt f.khan@fz-juelich.de}
\item Kup\'s\'c, Andrzej, Uppsala University, {\tt Andrzej.Kupsc@physics.uu.se}
\item Le\'sniak, Leonard, Henryk Niewodnicza\'nski Institute of Nuclear Physics,
Krak\'ow, {\tt Leonard.Lesniak@ifj.edu.pl}
\item Leupold, Stefan, Uppsala University, {\tt stefan.leupold@physics.uu.se}
\item Malaescu, Bogdan, CERN, CH--1211, Geneva 23, Switzerland, {\tt Bogdan.Malaescu@cern.ch}
\item Marianski, Bohdan, Narodowe Centrum Badan Jadrowych, {\tt bohdan@fuw.edu.pl}
\item Mascolo, Matteo, Universit\`a di Roma Tor Vergata, {\tt matteo.mascolo@roma2.infn.it}
\item Masjuan, Pere, Departamento de F\'isica Te\'orica y del Cosmos and CAFPE, Universidad de Granada, E-18071 Granada, Spain, {\tt masjuan@ugr.es}
\item Moskal, Pawel, Jagiellonian University, {\tt ufmoskal@if.uj.edu.pl}
\item Nyffeler, Andreas, Harish-Chandra Research Institute, Chhatnag Road, Jhusi,
Allahabad - 211019, India, {\tt nyffeler@hri.res.in}
\item Ostrick, Michael, Johannes Gutenberg University Mainz, Germany, {\tt ostrick@kph.uni-mainz.de}
\item Pel\'aez, Jos\'e R., Departamento de F\'{\i}sica Te\'orica II,
Facultad de CC. F\'{\i}sicas,  Universidad Complutense,
28040 Madrid, Spain, {\tt jrpelaez@fis.ucm.es}
\item Praszalowicz, Michal, Jagiellonian University, {\tt michal@if.uj.edu.pl}
\item Prencipe, Elizabetta, Johannes Gutenberg University Mainz, Germany, {\tt prencipe@kph.uni-mainz.de}
\item Przygoda, Witold, Jagiellonian University, {\tt witold.przygoda@uj.edu.pl}
\item Pszczel, Damian, National Centre for Nuclear Research, {\tt Damian.Pszczel@fuw.edu.pl}
\item Redmer, Christoph, Uppsala University, {\it current address:} Institut f\"ur Kernphysik, 
Johannes Gutenberg University Mainz, Germany, {\tt redmer@kph.uni-mainz.de}
\item Savinov, Vladimir, University of Pittsburgh, Dept. of Physics and Astronomy, 3941 O'Hara St., Pittsburgh, PA 15260, USA, {\tt vladimirsavinov@gmail.com}
\item Schadmand, Susan,	 Forschungszentrum J\"ulich, {\tt s.schadmand@fz-juelich.de}
\item Schneider, Sebastian P., Helmholtz-Institut f\"ur Strahlen- und Kernphysik (Theorie) and Bethe Center for Theoretical Physics, Universit\"at Bonn, Germany, {\tt schneider@hiskp.uni-bonn.de}
\item Stefanis, Nikolaos G., Institut f\"{u}r Theoretische Physik II, Ruhr-Universit\"{a}t Bochum,
D-44780 Bochum, Germany, {\tt stefanis@tp2.ruhr-uni-bochum.de}
\item Stepaniak, Joanna, National Centre for Nuclear Research (NCBJ) Warsaw/Swierk, {\tt Joanna.Stepaniak@fuw.edu.pl}
\item Stoffer, Peter, Albert Einstein Center for Fundamental Physics, Institute for Theoretical Physics, University of Bern, Sidlerstrasse 5, CH-3012 Bern, Switzerland, {\tt stoffer@itp.unibe.ch}
\item Terschl\"usen, Carla, Uppsala University, {\tt carla.terschluesen@physics.uu.se}
\item Uehara, Sadaharu, High Energy Accelerator Research Organization (KEK), Tsukuba, Japan, {\tt uehara@post.kek.jp}
\item Unverzagt, Marc, Institut f\"ur Kernphysik, Johannes Gutenberg University Mainz, Germany, {\tt unvemarc@kph.uni-mainz.de}
\item Wirzba, Andreas,	 Forschungszentrum J\"ulich, {\tt a.wirzba@fz-juelich.de}
\item Zdebik, Jaroslaw, Jagiellonian University, {\tt Jaroslaw.Zdebik@lnf.infn.it}
\item Zielinski, Marcin, Jagiellonian University, {\tt m.zielinski@uj.edu.pl}
\item Zupranski, Pawel, Narodowe Centrum Badan Jadrowych, {\tt zupran@fuw.edu.pl}
\end{itemize}
\end{flushleft}

\end{document}